\documentclass[12pt,preprint]{aastex}
\usepackage{epsf}
\usepackage{graphicx}

\newcommand{\sm}[1]{\mbox{{\scriptsize #1}}}
\newcommand{\simle} {\,{}^<_{\sim}\,}
\newcommand{\simge} {\,{}^>_{\sim}\,}

\newcommand{\be}{\begin{equation}}
\newcommand{\ee}{\end{equation}}
\newcommand{\bea}{\begin{eqnarray}}
\newcommand{\eea}{\end{eqnarray}}
\newcommand{\bdm}{\begin{displaymath}}
\newcommand{\edm}{\end{displaymath}}
\newcommand{\bef}{\begin{figure}}
\newcommand{\eef}{\end{figure}}
\newcommand{\befone}{
  \begin{figure*}
  \centering
  \begin{minipage}{\textwidth}
  }
\newcommand{\eefone}{\end{minipage}\end{figure*}}

\newcommand{\cm}{\mbox{cm}}
\newcommand{\m}{\mbox{m}}
\newcommand{\km}{\mbox{km}}
\newcommand{\AU}{\mbox{AU}}
\newcommand{\rad}{\mbox{rad}}
\newcommand{\g}{\mbox{g}}
\newcommand{\G}{\mbox{G}}
\newcommand{\Msol}{\mbox{$M_{\sun}$}}

\newcommand{\K}{\mbox{K}}
\newcommand{\yr}{\mbox{yr}}
\newcommand{\ys}{\mbox{yrs}}

\newcommand{\Ma}{{\cal M}}

\newcommand{\vx}{{\bf x}}

\newcommand{\dd}{{\mbox{d}}}

\def\eps@scaling{0.95}

\def\showone#1{
  \centering
  \leavevmode
  \epsfxsize=\eps@scaling\linewidth
  \epsfbox{#1.eps}
}

\def\epstwo@scaling{0.44}

\def\showtwo#1#2{
  \centering
  \leavevmode
  \epsfxsize=\epstwo@scaling\linewidth
  \epsfbox{#1.eps} 
  \epsfxsize=\epstwo@scaling\linewidth
  \epsfbox{#2.eps}
}

\def\showthree#1#2#3{
  \centering
  \leavevmode
  \epsfxsize=\eps@scaling\linewidth
  \epsfbox{#1.eps} 
  \epsfxsize=\eps@scaling\linewidth
  \epsfbox{#2.eps}
  \epsfxsize=\eps@scaling\linewidth
  \epsfbox{#3.eps}
}

\def\epstwo@scaling{0.44}

\def\showfour#1#2#3#4{
  \centering
  \leavevmode
  \epsfxsize=\epstwo@scaling\linewidth
  \epsfbox{#1.eps} \hfil
  \epsfxsize=\epstwo@scaling\linewidth
  \epsfbox{#2.eps} \hfil
  \epsfxsize=\epstwo@scaling\linewidth
  \epsfbox{#3.eps} \hfil
  \epsfxsize=\epstwo@scaling\linewidth
  \epsfbox{#4.eps}
}

\def\showsix#1#2#3#4#5#6{
  \centering
  \leavevmode
  \epsfxsize=\epstwo@scaling\linewidth
  \epsfbox{#1.eps} \hfil
  \epsfxsize=\epstwo@scaling\linewidth
  \epsfbox{#2.eps} \hfil
  \epsfxsize=\epstwo@scaling\linewidth
  \epsfbox{#3.eps} \hfil
  \epsfxsize=\epstwo@scaling\linewidth
  \epsfbox{#4.eps} \hfil
  \epsfxsize=\epstwo@scaling\linewidth
  \epsfbox{#5.eps} \hfil
  \epsfxsize=\epstwo@scaling\linewidth
  \epsfbox{#6.eps}
}

\shorttitle{Massive star formation}
\shortauthors{R.~Banerjee and R.E.~Pudritz}

\begin{document}

\title{Massive star formation via high accretion rates and early
  disk-driven outflows}

\author{Robi Banerjee\altaffilmark{1,2} and Ralph E. Pudritz\altaffilmark{2,3}}
\affil{$^1$Institute of Theoretical Astrophysics, University of
Heidelberg, Albert-Ueberle-Str. 2, 69120 Heidelberg, Germany \\
$^2$Department of Physics and Astronomy, McMaster University,
Hamilton, Ontario L8S 4M1, Canada \\ $^3$Origins Institute, McMaster
University, Arthur Bourns Bldg 241, Hamilton, Ontario L8S 4M1, Canada}

\begin{abstract}

We present an investigation of massive star formation that results
from the gravitational collapse of massive, magnetized molecular cloud
cores.  We investigate this by means of highly resolved, numerical
simulations of initial magnetized Bonnor-Ebert-Spheres that undergo
collapse and cooling.  By comparing three different cases - an
isothermal collapse, a collapse with radiative cooling, and a
magnetized collapse - we show that massive stars assemble quickly with
mass accretion rates exceeding $10^{-3} \, \Msol \, \yr^{-1}$. We
confirm that the mass accretion during the collapsing phase is much
more efficient than predicted by selfsimilar collapse solutions,
i.e. $\dot{M} \sim c^3/G$. We find that during protostellar assembly
the mass accretion reaches $20 - 100 \, c^3/G$.  Furthermore, we
determined the self-consistent structure of bipolar outflows that are
produced in our three dimensional magnetized collapse
simulations. These outflows produce cavities out of which radiation
pressure can be released, thereby reducing the limitations on the final
mass of massive stars formed by gravitational collapse.  Moreover, we
argue that the extraction of angular momentum by disk-threaded
magnetic fields and/or by the appearance of bars with spiral arms
significantly enhance the mass accretion rate, thereby helping the
massive protostar to assemble more quickly.

\end{abstract}


\section{Introduction}

The formation of massive stars is still a highly debated question.
Low mass stars accrete the bulk of their mass through their
circumstellar disks before nuclear burning turns on (eg. \cite{Shu87}).
Massive stars on the other hand, have Kelvin-Helmholtz time scales
that are smaller than the dynamical time so that young massive stars
start to burn their nuclear fuels while still accreting the
surrounding gas \citep[e.g.,][and references herein]{Yorke02b,
Yorke04}.  Infall, or flow of gas through a surrounding disk,
therefore faces a major obstacle in the form of the radiative
pressure that such massive stars will produce as they are still
forming.

Early spherical accretion
models suggest that the resulting radiation pressure could limit the
final mass of the star to $\sim 40 \, \Msol$ if the accretion rate is
not high enough ($\simge 10^{-3} \, \Msol \, \yr^{-1}$) \citep{Kahn74,
Wolfire87}. More recent two dimensional simulations by \citet{Yorke02}
showed that the limitations on the final mass of the massive star can
be relaxed if accretion through the protostellar disk is included in
models of massive star formation. Still, even this result yields mass
limits of $\sim 43 \, \Msol$. \citet{Krumholz05} argued that  
the effects of radiation pressure are limited by  
the escape of radiation through an outflow cavity.  
These results are were based on simulations using
Monte-Carlo-diffusion radiative transfer models where the outflow
cavities are parameterized by varying opening angles.

Alternatively, massive stars could form through coalescence of
intermediate mass stars~\citep[][]{Bonnell98}. This formation process
would be starkly different from the formation of low mass stars which
assemble quickly through accretion of the molecular gas. So far,
observations of the intermediate state of massive star formation are
rare and difficult to obtain. Nevertheless, a few massive objects
which show evidence for an ongoing accretion process are known by now
\citep[][]{Chini04, Patel05, Chini06, Beltran06}. 

We have developed detailed 3D collapse simulations of magnetized, star
forming cores and have shown that outflows and gravitational collapse
are inseparably linked.  In fact, we found in earlier work on the
assembly of low mass stars \citep[e.g.,][]{Banerjee06} that outflow
cavities are carved out of the collapsing envelope very early - much
before even a solar mass of material has been assembled into the star.
This important early outflow activity we suggested, could play a key
role in actually opening up a channel to radiation pressure in the
more massive systems.  The outflow channel is not imposed by an
external model but arises as a natural consequence of the generation
of an outflow during the  
formation of a magnetized disk.  

Our first study of massive star formation was in the context of the
turbulent fragmentation picture.  Specifically, we followed the
formation of massive stars in supersonic turbulence by using a 3D,
adaptive mesh (AMR) simulation and showed that filamentary accretion
plays an important role in focusing the collapse of material onto a
massive young star and its disk~\citep[][henceforth
BPA06]{Banerjee06b}.  We included a full list of coolants and found
that the central regions of the collapse behaved somewhat like
Bonnor-Ebert spheres.  We did not have MHD at work in those
calculations.

The present paper follows up on these previous efforts.  Instead of
trying to track the formation of a massive star in a very large
cluster simulation, we concentrate here on the collapse of one massive
central region chosen to mimic the massive collapse we found in BPA06.
The choice of an initial magnetized Bonnor-Ebert sphere is a 
reasonable starting point for such a focused-in study.  In particular,
we study the collapse of a massive molecular cloud core with a total
mass of $\sim 170 \, \Msol$, which we model as a sphere that is
initially in in pressure equilibrium with the surrounding medium
i.e. Bonnor-Ebert-Spheres (BE-Spheres).

We shall show that outflows are also 
an early aspect of the formation of massive 
stars.   We self-consistently determine the
initial scale and structure of outflow cavities that are carved by
magnetic tower flows that are driven by the forming protostellar disk.
In particular we study the influence of radiative cooling and magnetic
fields on the collapse of such cloud cores thereby discussing the
similarities and differences of these models.  We find that these
outflows are launched very early, much before the central stars have
even accumulated a solar mass worth of material.  This means that
even before radiation pressure becomes a factor in the formation of 
a massive star, the outflow has created the cavity out which the  
radiation pressure, when it eventually appears, can be released.  
Early outflows therefore, play a central role in the assembly
of such stars.

Our paper is organized as follows. In Sec.~\ref{sec:numerics} we
describe the numerical modelling and initial conditions of our
simulations. We discuss the accretion process and rates in
Sec.~\ref{sec:accretion}. In Sec.~\ref{sec:angular_momentum} we
discuss the angular momentum evolution and distribution for the
different models,and we discuss the influence of the early outflows in
Sec.~\ref{sec:outflows}. Finally, we conclude the results of this work
in Sec.~\ref{sec:conclusions}.

\section{Numerical modeling}
\label{sec:numerics}

For this study we follow the collapse of
Bonnor-Ebert-type~\citep[][]{Ebert55, Bonnor56} molecular cloud cores
using the AMR based FLASH code~\citep{FLASH00} to solve the
gravito-magneto-hydrodynamic equations (details on the
Bonnor-Ebert-Profile and of this setup can be found in
~\cite{Banerjee04}, henceforth BPH04). We study the collapse of three
different cases which all have the same initial density and
temperature distribution. The three cases are:
\begin{itemize}
\item The isothermal case (iso): Here, we assume an isothermal
  equation of state (adiabatic index, $\gamma = 1.0001$) throughout
  the calculation.
\item The hydro case (hydro): For this case we incorporate the cooling
  ability from molecular and dust radiation in the optical thin regime
  and use a radiation diffusion approximation in the optical thick
  regime. Details of our cooling treatment can be found in the appendix
  of BPA06.
\item The magnetized case (mag): Here, we apply an initial magnetic
  field parallel to the rotation axis ($z$-axis) and use the same
  cooling as in the pure hydro case.
\end{itemize}

The three different cases presented in this study, have the same
initial values for the total mass of the cloud core, $M = 168 \,
\Msol$, its radius, $R = 3.34\times 10^5 \, \AU = 5\times 10^{18} \,
\cm$, and the BE-core density, $\rho_0 = 3.35\times 10^{-21} \g \,
\cm^{-3}$. We also enhance the density of all cores by 10\% and add a
10\% $m =2$ density perturbation. The initial free fall time
associated with the core region is $t_{\sm{ff}} =
\sqrt{3\pi/32\,G\,\rho_0} = 3.63 \times 10^{13} \, \sec = 1.1\times
10^6 \, \ys$. We also give the cloud cores a slight spin with a
constant angular velocity. The two non-magnetic simulations start with
an angular velocity of $\Omega = 5.5\times 10^{-15} \, \rad \,
\sec^{-1}$ whereas the magnetized case rotates with an angular
velocity of $\Omega = 1.1\times 10^{-14} \, \rad \, \sec^{-1}$. We
increase the initial rotation by a factor of two in the magnetized
case because magnetic braking spins down the sphere substantially
prior to the collapse~\citep[see][]{Banerjee06}.  These initial
conditions are equivalent to an $\Omega\,t_{\sm{ff}}$ of $0.2$ and
$0.4$ for the non-magnetized and magnetized cases, respectively. In
BPH04 we showed that the former case results in a ring which fragments
into a binary system if only molecular cooling is considered. The
initial isothermal sound speed, $c_{\sm{iso}}$, is in all cases $0.408
\, \km \, \sec^{-1}$. 

Initially, the magnetic field in the magnetized case is given only a
component parallel to the rotation axis ($z$-axis) where $B_z$ varies
between $0.33 \, \mu\G$ and $1.36 \, \mu\G$ corresponding to a
constant $\beta \equiv p/(B^2/8\pi) = 76$ in the equatorial
plane. This gives a mass-to-flux ratio of $M/\Phi_B = 3.2/\sqrt{G}$
which is highly supercritial (the critial value is $\sim
0.12/\sqrt{G}$). Nevertheless the magnetic field becomes dynamically
important during the collapse and will be able to drive outflows and
jets as we will see later.  This choice of a supercritical core is in
agreement with recent work by \citet{Tilley05} which studied the
formation of turbulent fragmentation and the IMF in magnetized
turbulent clumps.  In that paper, it was demonstrated that
substantially supercritical clumps did a far better job in reproducing
the observations of the core mass function in cluster forming
environments, than did clumps that were nearer to magnetically
critical values.  One way that this can easily occur in a turbulent
environment is for shocks to rapidly assemble cores by flows along
field lines.

The choice of our initial core mass is taken from recent
observations of cores in regions of high mass star formation 
\citep[e.g.,][]{Reid06a}. The actual physical
values we use arise in this context arise
from the consideration of a BE
sphere with a core density of $10^3 \, \cm^{-3}$ ($3.35\times 10^{-21}
\, \g \, \cm^{-3}$) at $20 \, K$ which is a typical value of such molecular
cores. As we will see later, the mass accretion rate depends only
weakly on the initial mass of the core.

\section{High accretion rates during the collapse phase}
\label{sec:accretion}

\bef
\showtwo{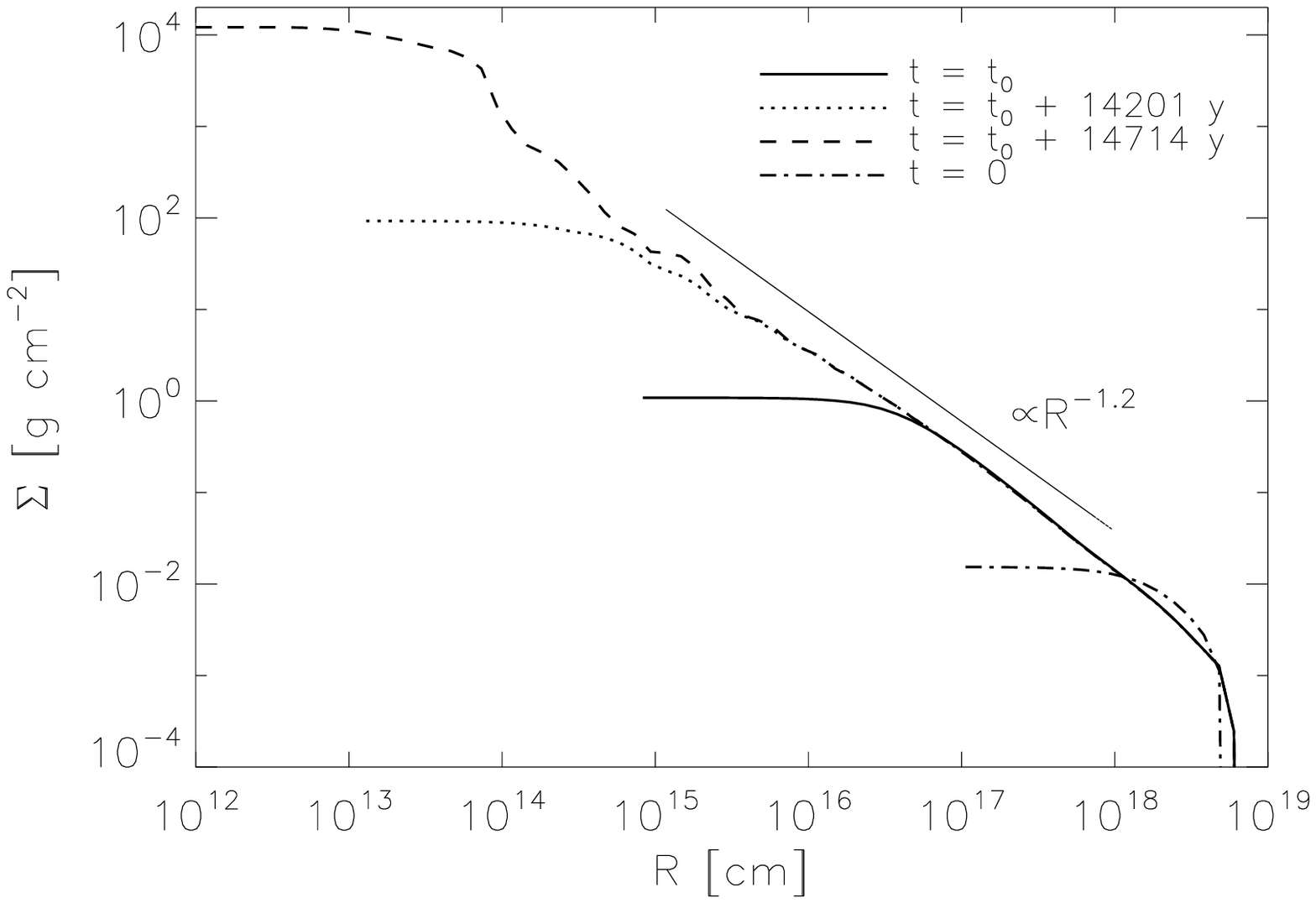}
	{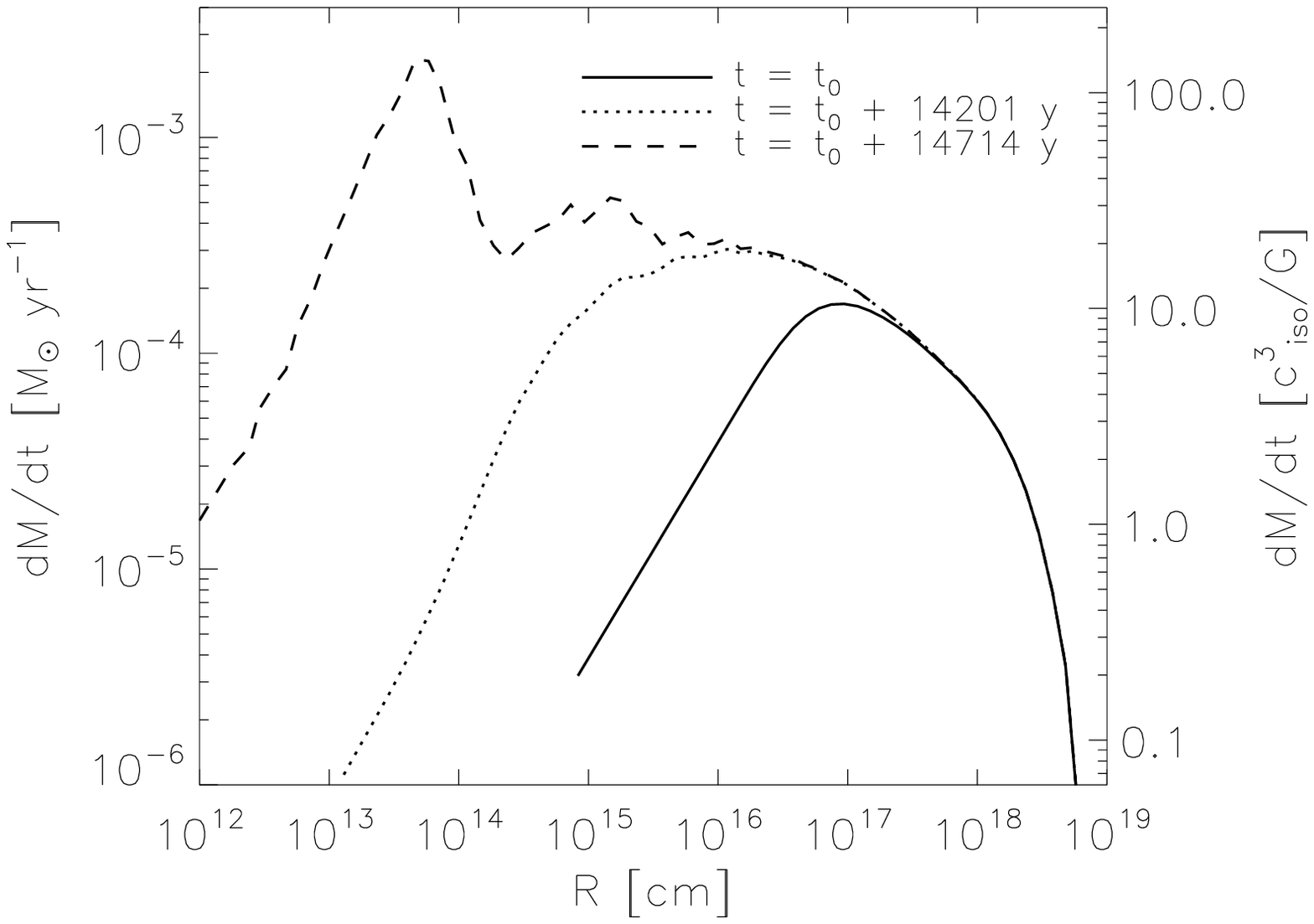}
\caption{Shows the time evolution of the column density and mass
  accretion in the magnetized case.
%
% nfs = [129, 156, 320, 0]
% values at file 129:
%    t_0 = 1.6023356e+14 sec = 5077495.0 years
%    rho_core = 2.44d-17 g cm^-3
%    tff = 13468.6 years
%
We assume the beginning of the collapsing phase at $t = t_0$ when the
column density reaches $\Sigma_{\sm{core}} \approx 1 \, \g \,
\cm^{-2}$ and the pressure is $P/k_B \approx 10^8 \, \K \, \cm^{-3}$
(see Fig.~\ref{fig:evol_press}). At $t = t_0$ the mass density in the
core is $\rho_{\sm{core}} = 2.44\times 10^{-17} \, \g \, \cm^{-3}$
corresponding to a free fall time of $t_{\sm{ff}} = 1.35\times 10^4 \,
\ys$. At $t = t_0 + 1.1 \, t_{\sm{ff}}$ the core density reached
already $10^4 \, \g \, \cm^{-2}$. The initial profile of our
simulation is marked by $t = 0$.}
\label{fig:evol_dens}
\eef

\bef
\showtwo{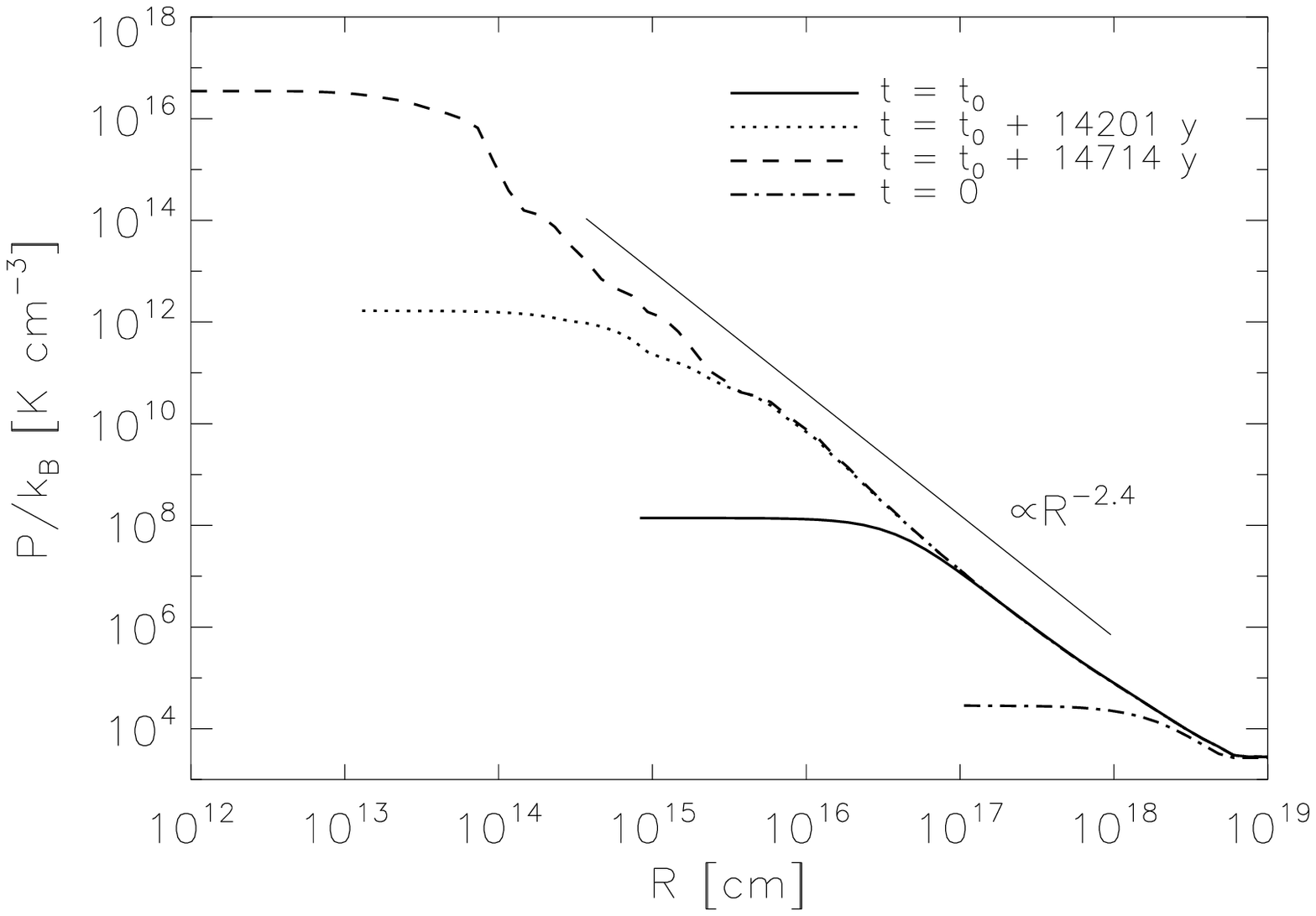}
	{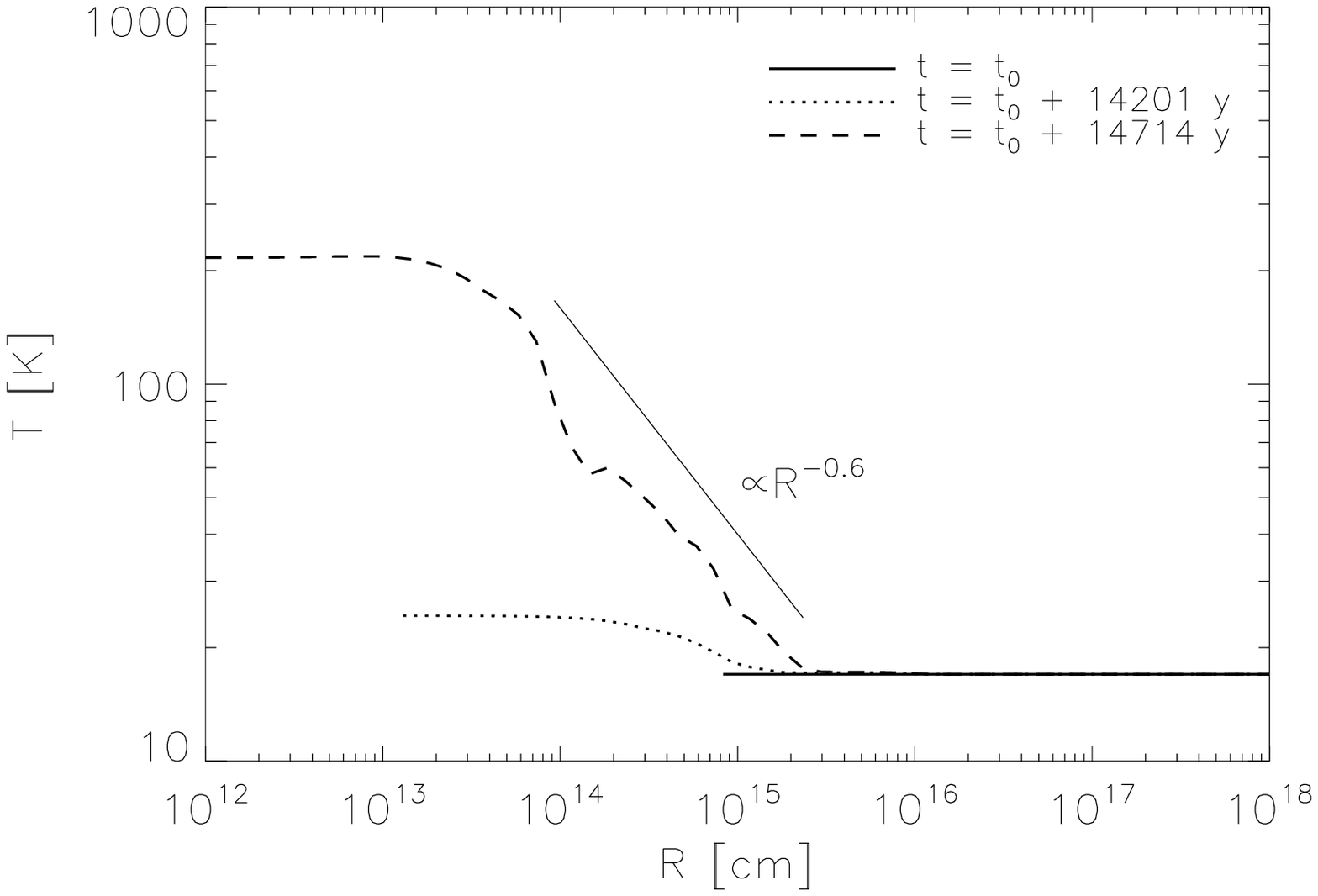}
\caption{Shows the time evolution of the pressure and temperature in
  the magnetized case. See Fig.~\ref{fig:evol_dens} for the
  corresponding column densities and label explanations. The pressure
  scales as $P \propto \Sigma^2$ and the temperature develops a
  $R^{-0.6}$ profile within the warm, dense region.
%
% nfs = [129, 156, 320, 0]
}
\label{fig:evol_press}
\eef

The collapse of spherical cloud cores has been studied by many
authors, both analytically and numerically \citep[see
e.g.,][]{Larson69, Penston69, Shu77, Hunter77, Whitworth85, Foster93,
Hennebelle03, Banerjee04}. In the case of an initial singular
isothermal sphere (SIS) there exists an elegant self-similar solution
to this problem wherein an expansion wave travels from the (singular)
center outward with the speed of sound, thereby initiating an inside-out
collapse~\citep{Shu77}. However, as pointed out by \citet{Whitworth96} a
SIS configuration is unnatural because -- among other difficulties --
its collapse can not produce binaries \citep[see
also][]{Pringle89}. The collapse of non-singular (ie Bonner
-Ebert ) cores proceed
differently than singular spheres and have distinguishable
implications. Firstly -- as demonstrated by many authors
\citep[e.g.,][]{Larson69, Penston69, Foster93, Hennebelle03,
Banerjee04} -- the collapse proceeds from the {\em outside-in} rather than
from the inside-out and the density maintains a flat profile at the core
center, where the core size is of the order of the local Jeans length
at every epoch~\citep[see also][for a summary of analytic
solutions]{Whitworth85}. Secondly, the radial distribution of the infall
velocity peaks at the edge of the flat density core and falls off
quickly towards the center (e.g. see Fig.~3 in BPH04). The
velocity also becomes supersonic as the core density increases and the
size of the flat region shrinks. Recent observations of a pre-Class 0
object show that the collapse proceeds supersonically, strongly
supporting a Larson-Penston-type collapse rather than an expansion
wave-type collapse \citep{Furuya06}.  Thirdly -- and of great importance
-- the mass accretion in the non-selfsimilar case is much higher than
predicted from the selfsimilar collapse of a singular isothermal
sphere. We find that the mass accretion in the early phase of the
collapse is
\be 
   \dot{M} \approx 20 - 100 \, \frac{c^3}{G}
\label{eq:mdot}
\ee 
($c$ is the isothermal sound speed and $G$ is Newton's
gravitational constant). Note the selfsimilar SIS collapse gives a
mass accretion of only $0.96 \, c^3 / G$. Typical values of the sound
speed in cold cloud cores ($T \sim 20 \, \K$) are of the order of a
few $10^2 \, \m \, \sec^{-1}$ which gives $c^3/G \sim 10^{-6} -
10^{-5} \Msol \, \yr^{-1}$. Our result from numerical simulations are
in agreement with the early analytic results of \citet{Larson69} and
\cite{Penston69} \citep[see also][]{Hunter77, Whitworth85}. 

The remarkable point of this result is that the high accretion rates
are achieved even without initial turbulence and during the isothermal
phase of the collapse. The main reason for the high accretion rate in
this idealized case is the supersonic infall velocity, $v_r$, close to
the peak density. Even moderate Mach numbers, $\Ma$, of $2 - 3$ (The
Larson asymptotic Mach number is $\Ma \sim 3$) enhances the mass
accretion rates relative to $c^3/G$ because 
\be
c^3/G \to v_r^3/G = \Ma^3 \, c^3/G
\label{eq:accretion}
\ee
in the supersonic limit. Additionally, the core is continually
embedded in a high pressure environment.

One-dimensional numerical studies of the long term 
evolution (i.e. beyond one dynamical time) of
a collapsing BE-Sphere by~\citet{Foster93} showed that the mass
accretion is not constant but decreases with time after reaching a
peak value. Beyond this peak, the remaining gas in the envelope drains out 
with an ever smaller accretion rate over time. This situation could be
different for the collapse of realistic cloud cores, which are certainly not
isolated objects but are instead surrounded by a clumpy media and
attached to filaments wherein accretion
might sustain high rates for a longer time.

Using the magnetized simulation as the fiducial case, we show the
evolution of the column density and corresponding mass accretion rates
in Fig.~\ref{fig:evol_dens} and in Fig.~\ref{fig:evol_press} the
evolution of the pressure and temperature as functions of the disk
radius~\footnote{All quantities which are functions of the disk
radius, $R$, are azimuthally averaged and density weighted, e.g. $f(R)
= (2\pi \, \Sigma(R))^{-1} \, \int \dd z \, \dd\phi \, \rho(\vx) \,
f(\vx)$, where $\Sigma(R) = (2\pi)^{-1} \, \int \dd\phi \, \dd z \,
\rho(\vx)$ is the column density.}. When the density approaches
$\Sigma \sim 1 \, \g \, \cm^{-2}$ (we denote the corresponding time as
$t_0$) the mass accretion rate is already $\sim 10^{-4} \, \Msol \,
\yr^{-1}$ and the surrounding pressure is $10^8 \, \K \, \cm^{-3} \,
k_B$. These results are in agreement with the model of~\citet{McKee02,
McKee03} who showed that this high pressurized, compact cloud cores
accrete gas with accretion rates as high as $10^{-3} \, \Msol \,
\yr^{-1}$. We find that accretion rates of this order ($\sim 10^{-3}
\, \Msol \, \yr^{-1}$) are reached within only $\sim 14.7\times 10^3
\, \ys$ which corresponds to $1.1$ dynamical times. The core column
density after one dynamical time is $\sim 10^4 \, \g \, \cm^{-2}$
while the pressure achieves a value $P/k_B \approx 3.5 \times 10^{16}
\, \K \, \cm^{-3}$. As long as the core remains isothermal (the most
efficient cooling regime), the pressure scales with the column density
as $P \propto \Sigma^2$ (i.e. the pressure profile is close to
$R^{-2.4}$) and slightly steeper in the inefficient cooling regime
where the temperature rises during the collapse (see right panel of
Fig.~\ref{fig:evol_press}). The continuously high external pressure
and the supersonic infall velocity maintains the high mass
accretion. We also show the mass accretion scaled to the quantity
$c_{\sm{iso}}^3/G$ ($c_{\sm{iso}}$ is the {\em initial} isothermal
sound speed) in Fig.~\ref{fig:evol_dens}. From this one can see that
accretion is much more efficient (by a factor of $\sim 100$) than
expected from the collapse of a SIS.

These results show that high mass accretion rates are a natural result
of the early collapse phase of (non-singular) collapsing cloud cores
with flat-topped density profiles. Turbulent driving for the rapid
assembly of massive stars is not a necessary ingredient, although it does
further enhance the accretion rate as shown in analytic
models~\citep{McKee02, McKee03}. The enhancement is even larger
in fully
realized 3D turbulence, due to filamentary accretion (see e.g., BPA06).

\bef
\showtwo{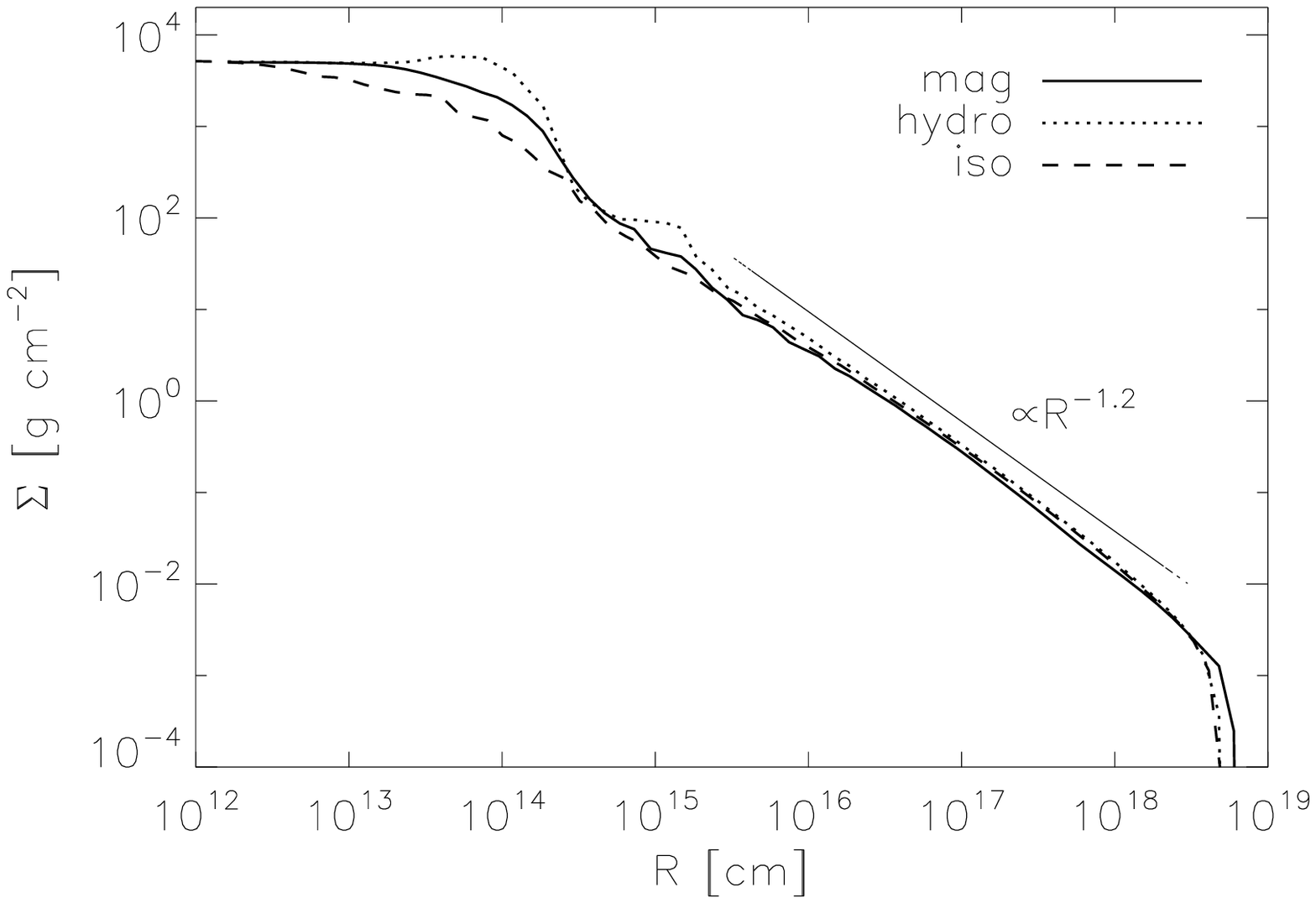}
	{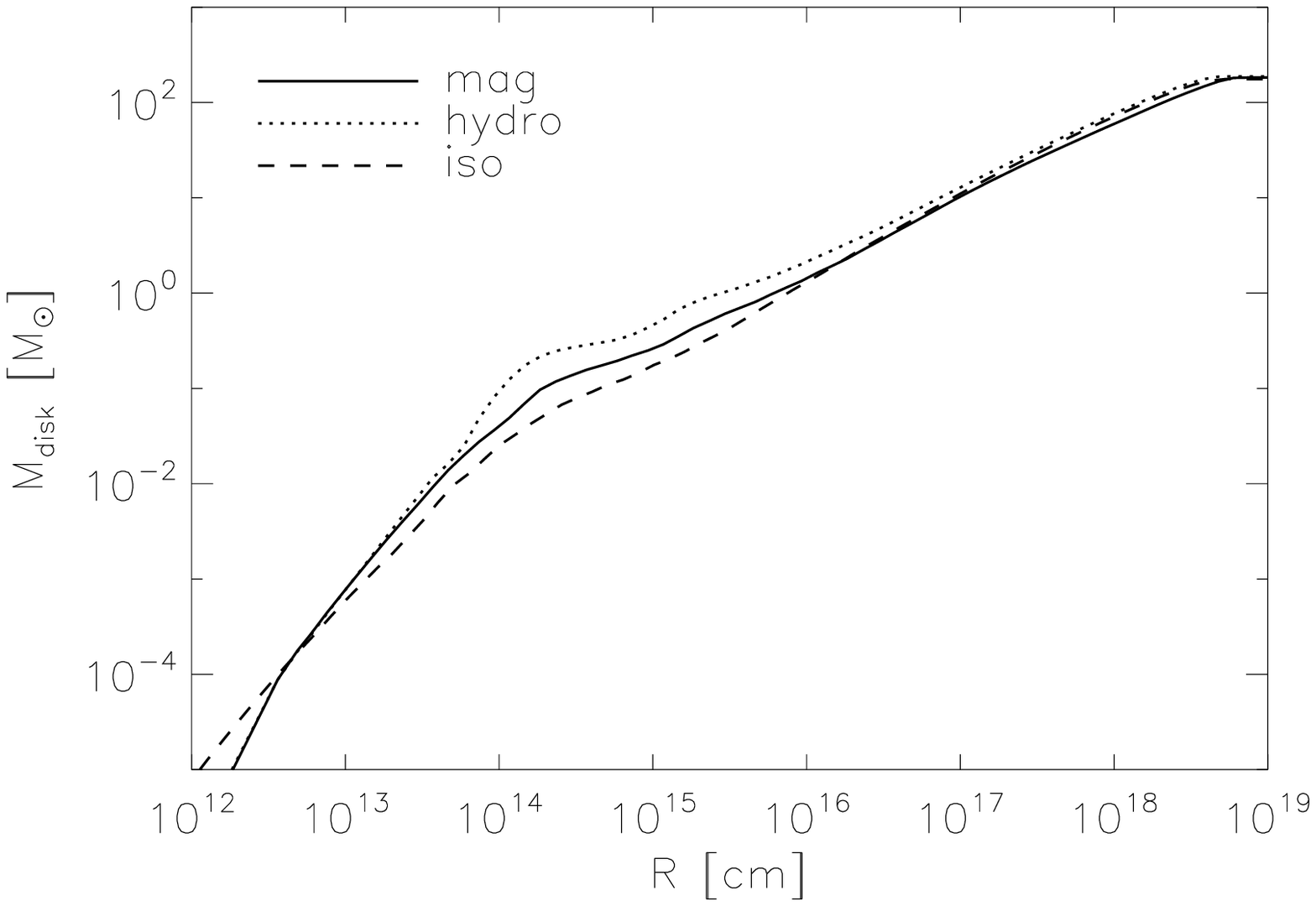}
\caption{Comparison of the column density profiles and disk masses
  for three different cases, hydromagnetic (mag), pure hydro (hydro),
  and pure isothermal (iso) simulation at the time when they reached
  the same core column density $\Sigma_{\sm{core}} \sim 5\times 10^3
  \, \g \, \cm^{-2}$.}
%
% dir = ['BE_mag_dust/dat/', 'BE_dust/dat/', 'BE_ISO_rot_v1/dat/']
% nfs = [240, 214, 169]
%
\label{fig:column_density}
\eef

\bef
\showtwo{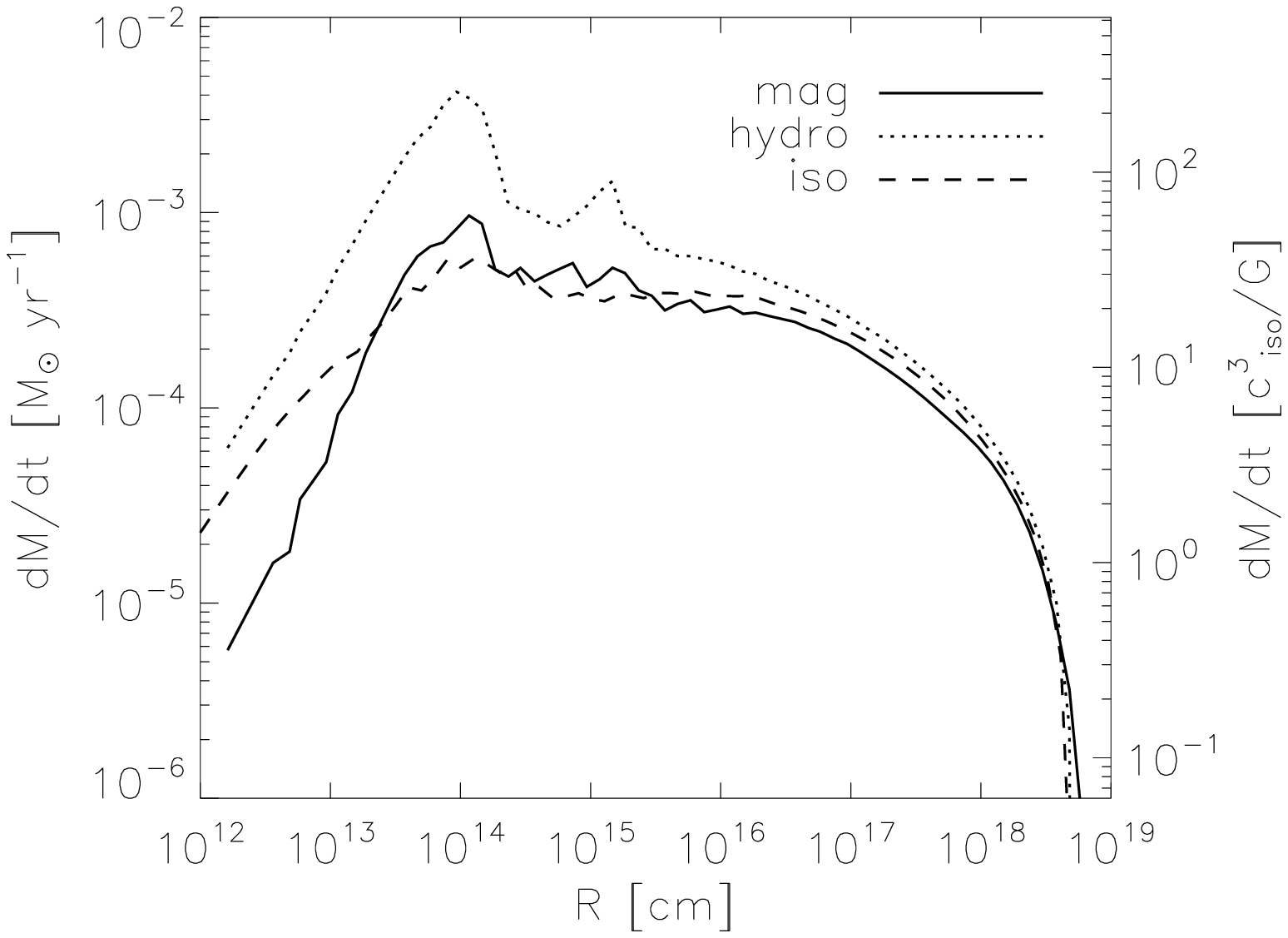}
        {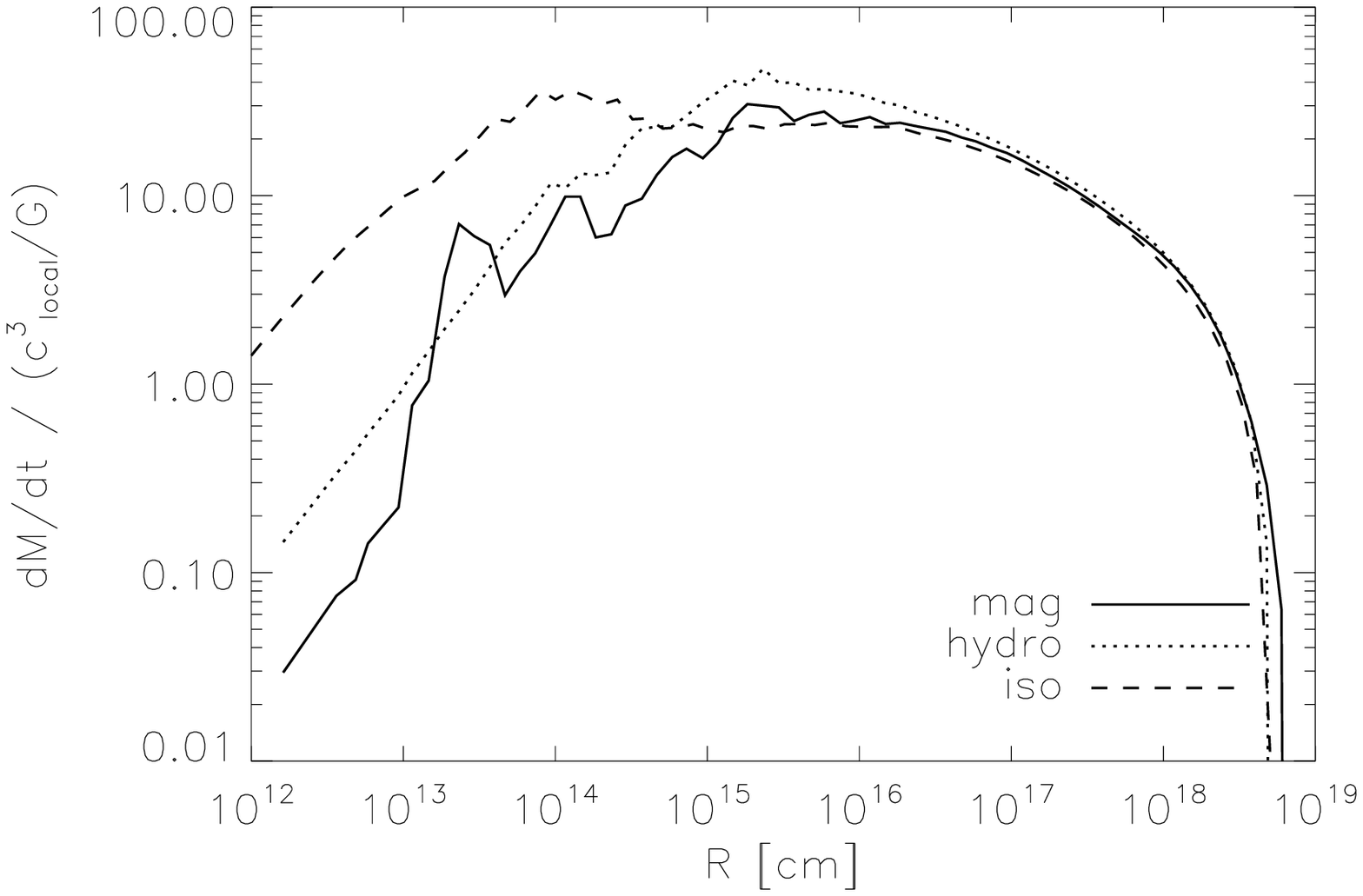}
\caption{Comparison of the mass accretion rate for the three different
  cases, hydromagnetic (mag), pure hydro (hydro), and pure isothermal
  (iso) simulation at the time they reached the same core density
  (see Fig.~\ref{fig:column_density}). The left panel shows the mass accretion
  in units of $\Msol \, \yr^{-1}$ whereas the right panel shows the
  same quantity compared to $c_{\sm{local}}^3/G$, where
  $c_{\sm{local}}$ is {\em local} sound speed.}
\label{fig:mdots}
\eef

\bef
\showtwo{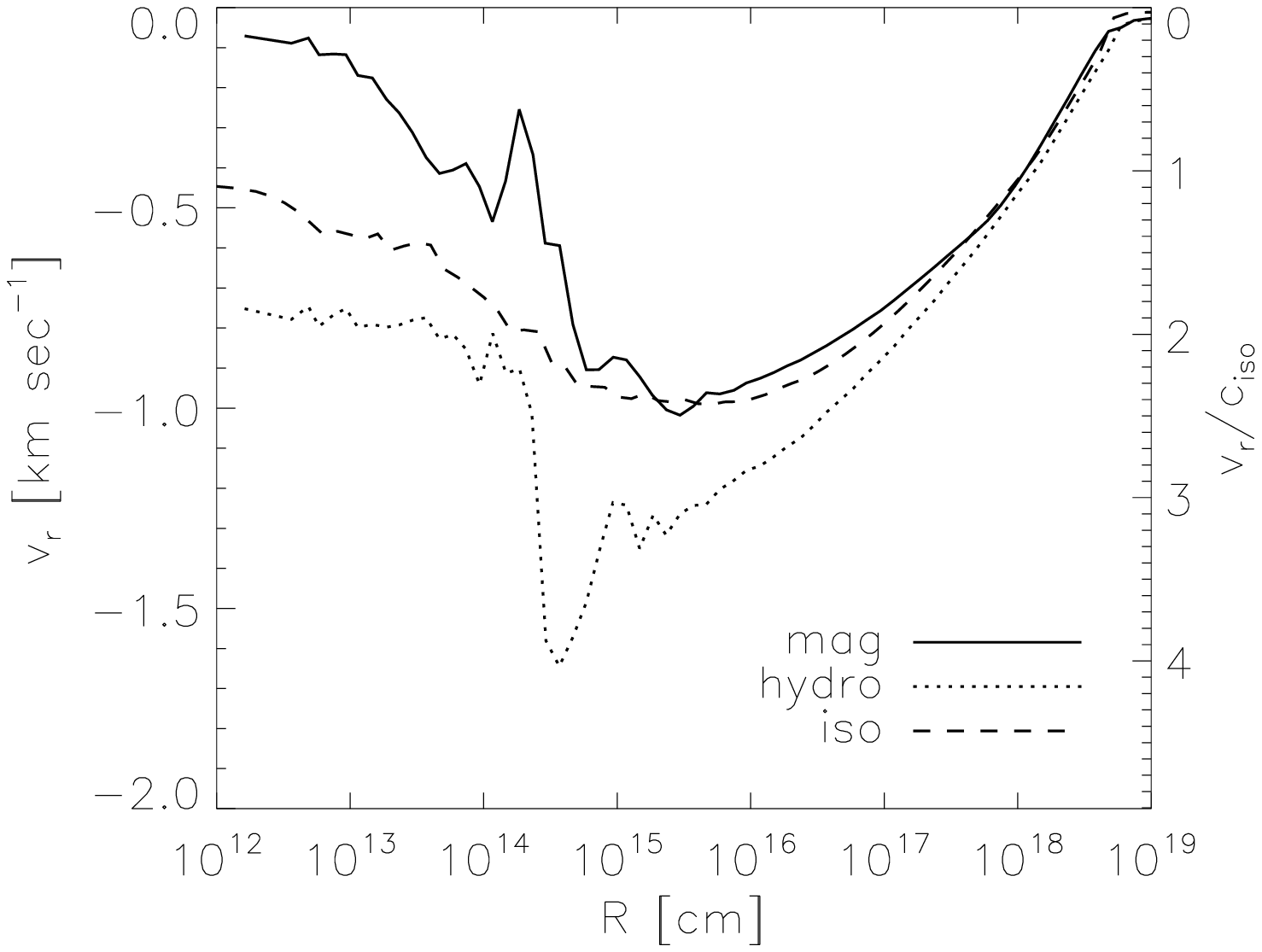}
	{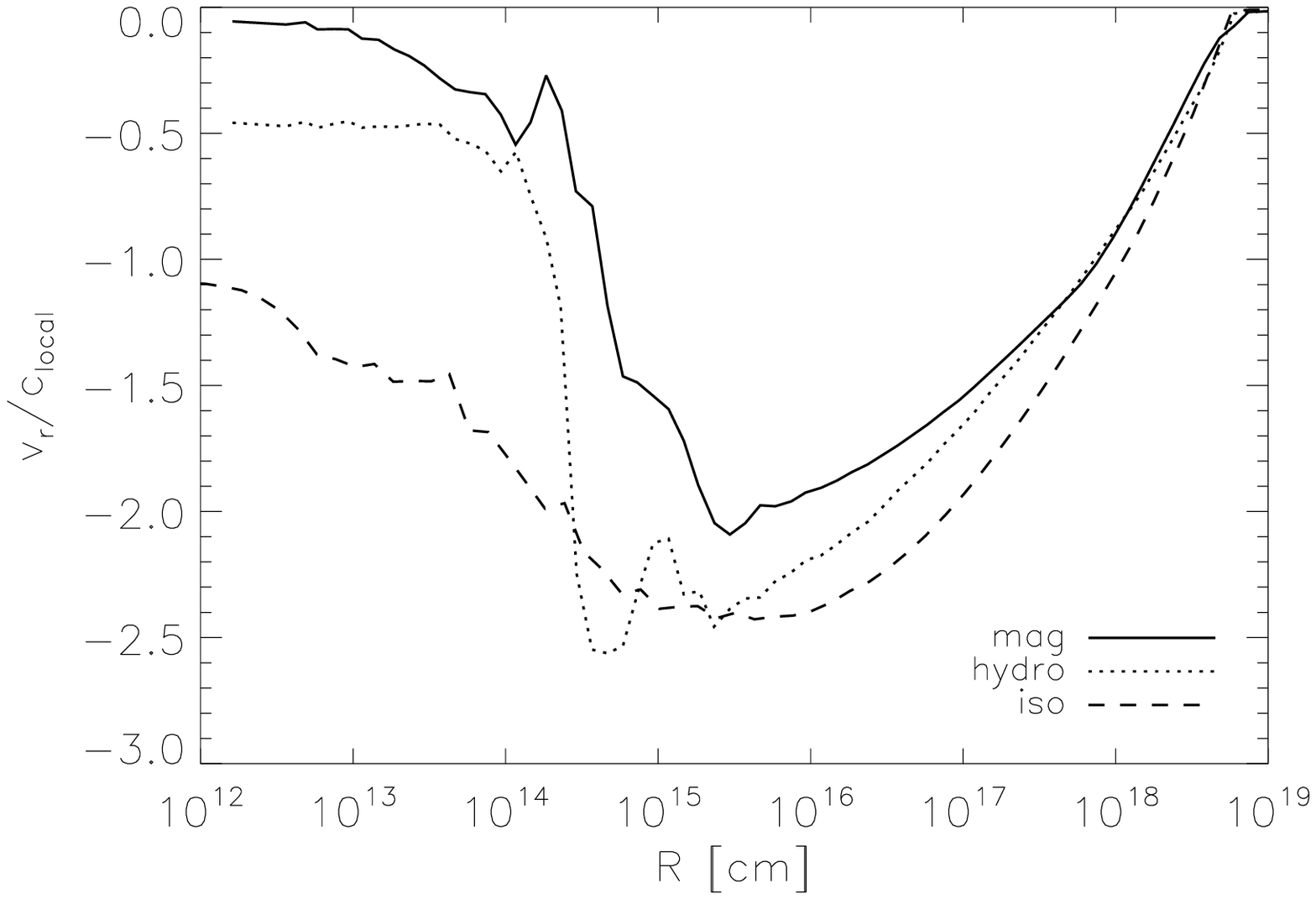}
\caption{Comparison of the radial infall velocity for the three
  different cases, hydromagnetic (mag), pure hydro (hydro), and pure
  isothermal (iso) simulation at the time they reached the same core
  density (see Fig.~\ref{fig:column_density}). The left panel shows
  the infall velocity in km/sec and compared to the initial
  isothermal sound speed, $c_{\sm{iso}}$, and the right panel shows
  the same quantity measured with the {\em local} sound speed,
  $c_{\sm{local}}$.}
\label{fig:vrad}
\eef

For comparison, we summarize the collapse phase of the three different
cases (mag, hydro, iso) in the Figs.~\ref{fig:column_density} --
\ref{fig:vrad} at the time when the column density reaches $5\times
10^3 \, \g \, \cm^{-2}$. Common features can be seen mainly in the
envelope of the collapsing cores. All three cases end up with a column
density profile which is close to $\propto R^{-1.2}$ in the disk
envelope (see left panel of Fig.~\ref{fig:column_density}). In the
inner part of the disk ($\simle 100 \, \AU$) the influence of the
(inefficient) cooling processes and magnetic fields become
visible. The non-isothermal cases develop a steeper density profile as
infalling material tends to pile up at the shock regions at the edge
of the warm core. These shock regions appear at a radial distance of
$\sim 20 \, \AU$ because the core becomes opaque inside this
radius~(BPA06). The difference with the isothermal case
inside $\simle 100 \, \AU$ can also be seen in the mass distribution
of the disk which we show in the right panel of
Fig.~\ref{fig:column_density}. More material accumulates in the hydro
and magnetized cases at the time when all three cases reached a peak
column density of $5\times 10^3 \, \g \, \cm^{-2}$.

We show the comparison of the mass accretion rates of the three
different cases in Fig.~\ref{fig:mdots}. The right panel gives the
mass accretion in terms of solar masses per year and compared to the
initial isothermal sound speed, whereas the right panel shows this
value scaled to the local sound speed quantity
$c_{\sm{local}}^3/G$. In physical terms (i.e. $\Msol/\yr$) the hydro
case is most advanced and accretes gas with several $10^{-3} \, \Msol
\, \yr^{-1}$ corresponding to $\sim 300 \, c_{\sm{iso}}^3/G$. Then
again, the accretion rates of all three cases are very similar when
compared to local quantity $c_{\sm{local}}^3/G$ ($\sim 40 \,
c_{\sm{local}}^3/G$). This shows, again, that the estimate for the mass
accretion in the supersonic regime is sensitive to the Mach number and
the local sound speed. To complete this picture, we show the infall
velocity in Fig.~\ref{fig:vrad} measured in $\km/\sec$ and
$c_{\sm{iso}}$ (left panel) and in terms of the local sound speed
(right panel). The high infall velocity (and therefore the high mass
accretion) in the hydro case is due to the warm core region which has
a high sound speed. The local Mach number of all three cases differs only
within a range $2 - 2.5$ whereas the isothermal Mach number ranges
from $2.5 - 4$ between them.

We also point out that the mass accretion during this early stage of
collapse does not strongly depend on the initial mass of the cloud
core (the good estimate of Eq.~\ref{eq:accretion} does not have a mass
dependence). In BPH04 and \citet{Banerjee06} we showed that a $\sim 2
\Msol$ cloud core reaches also mass accretion rates of $10^{-4} -
10^{-3} \Msol/\yr$.

In summary, we find that the typical infall velocity of a collapsing
cloud core is $2 - 3 \, c_{\sm{local}}$ which recovers earlier results
studying the isothermal collapse phase \citep[e.g.,][]{Larson69,
  Foster93}. The supersonic infall velocity is the main reason for the
high mass accretion of collapsing cloud cores. Furthermore, the
increased local sound speed due to thermal effects in the warm core
region enhances the mass accretion in the post-isothermal regime.

\section{Angular momentum extraction by bars and magnetic fields}
\label{sec:angular_momentum}

\bef
\showtwo{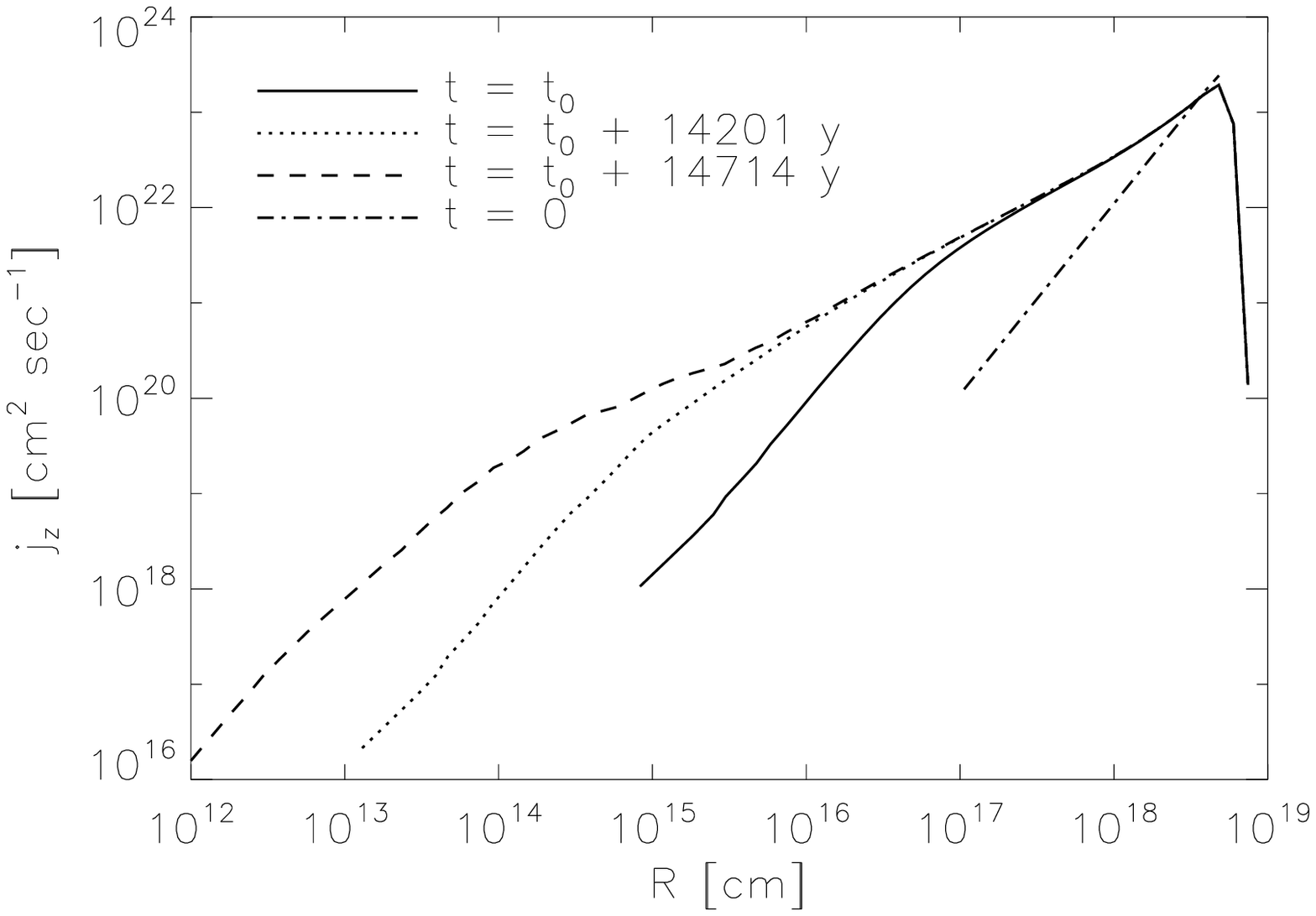}
	{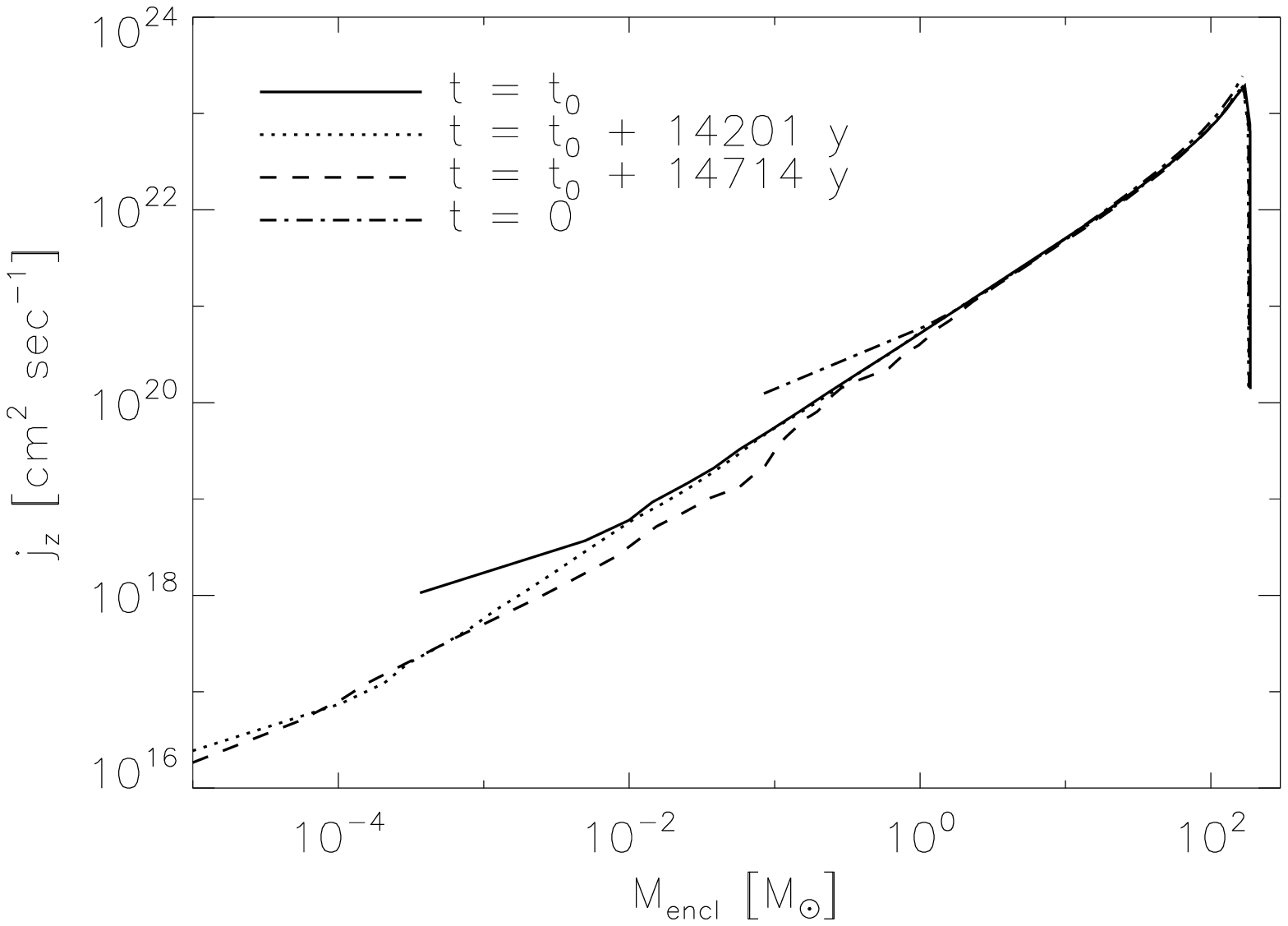}
\caption{Shows the time evolution specific angular momentum as a
  function of radius and enclosed mass in the magnetized case. See
  Fig.~\ref{fig:evol_dens} for the corresponding column densities and
  label explanations.
%
% nfs = [129, 156, 320, 0]
}
\label{fig:evol_jz}
\eef

Most stars, if not all, rotate.  The origin of stellar
angular momentum is in itself an important problem in 
star formation theory.  It has a natural explanation in the
turbulent fragmentation picture, where it arises from the oblique
shocks that create the dense cores (eg. Tilley \& Pudritz 2004).
Strict conservation of the specific
angular momentum in a core would inhibits accretion through the protostellar
disk whenever the infalling gas hits its centrifugal barrier. The
issue of redistributing or extracting the angular momentum within the
protostellar disk is important for understanding the assembly of (massive)
stars. 

In Fig.~\ref{fig:evol_jz} we show the evolution of the specific
angular momentum in the magnetized case as a function of radius (left
panel) and enclosed mass in the disk (right panel). Infalling, high
angular momentum gas increases the specific angular momentum at small
radial distances with time. Apart from the advection of angular
momenta due to the accreting gas, there is additionally, a resorting of
disk angular momentum operating due to turbulence. 
This can be seen from the right panel of
Fig.~\ref{fig:evol_jz} where we plot the specific angular momentum as a
function of the enclosed mass.  This is a useful thing to do 
because this quantity is constant with time if angular
momentum is {\em not} redistributed within the disk \citep[see
e.g.,][]{Abel02}. For the magnetic case shown in
Fig.~\ref{fig:evol_jz} the angular momentum is {\em decreasing} with
time within the disk (inside $M_{\sm{disk}} \simle 1 \, \Msol$). This
decrease of angular momentum at a given mass shell allows for
continuing efficient accretion through the disk plane. In particular,
we find that the specific angular momentum changes by an amount of
$\Delta j_z \sim 5.3\times 10^{19} \, \cm^2 \, \sec^{-1}$ (i.e. by a
factor of 2.8) at a mass shell of $0.1 \,\Msol$ within $14.7\times
10^3 \, \ys$. In the magnetized case the angular momentum is extracted
by magnetic braking, magnetic torque, the launching of an outflow, and
by the development of a bar in the disk. Depending on the field
geometry and strength magnetic braking by a poloidally dominated
magnetic field can be an efficient way to extract angular
momentum~\citep[][]{Mouschovias79, Mouschovias80}. An additional
toroidal component of the magnetic field gives rise to a magnetic
torque \citep[e.g.,][]{Konigl00} which in the case of rotationally
generated toroidal field (backwards bend field) extracts angular
momentum from the protodisk. In \citet{Banerjee06} we showed that the
magnetic torque can be a substantial fraction (10\% -- 50\%) of the
angular momentum flux from the accreting gas.

\bef
\showone{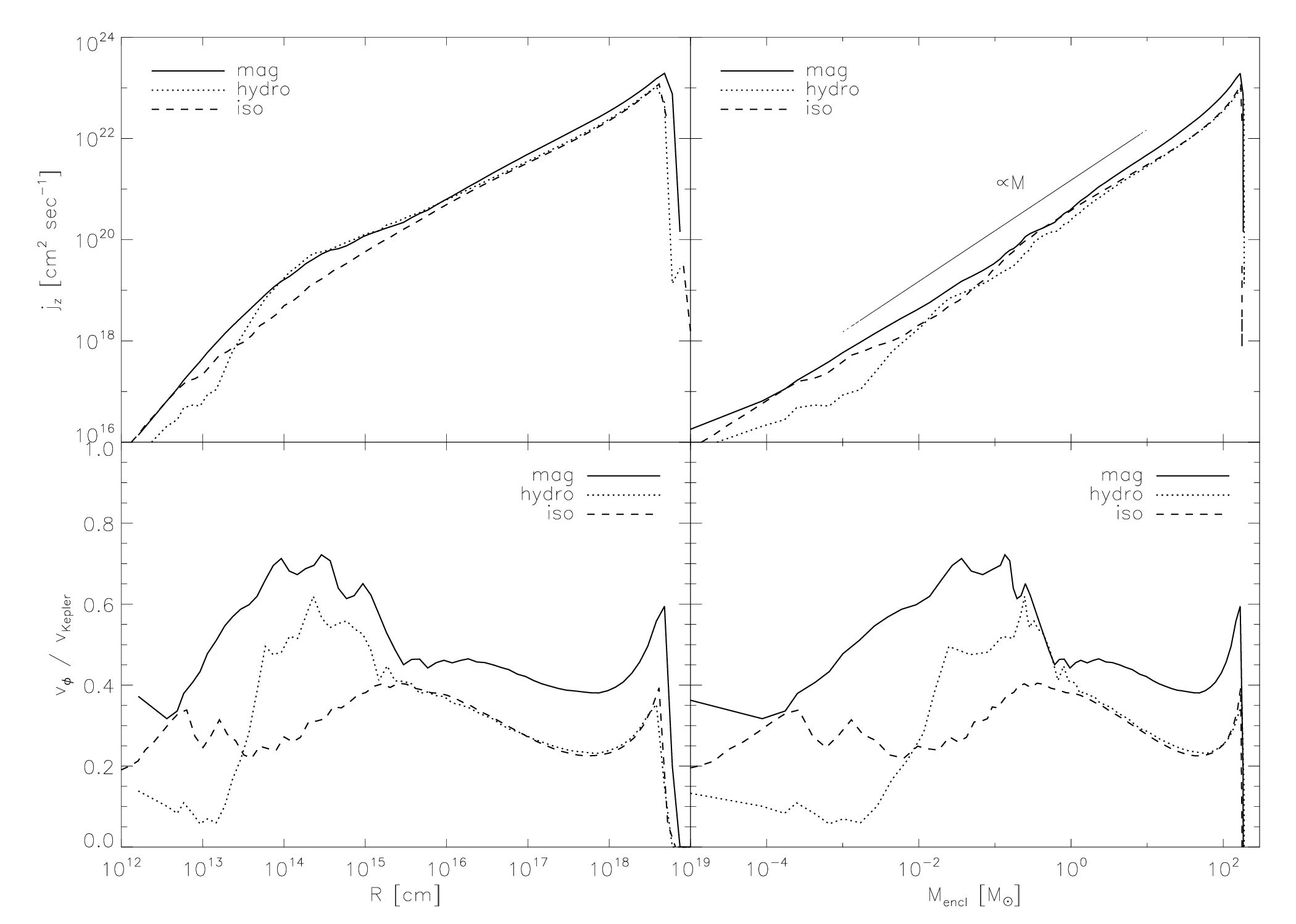}
\caption{Comparison of the specific angular momenta and rotational
  velocity for the three different cases, hydromagnetic (mag), pure
  hydro (hydro), and pure isothermal (iso) simulation at the time they
  reached the same core density of $5\times 10^3 \, \g \, \cm^{-2}$
  (see Fig.~\ref{fig:column_density}). The left panels show this
  quantities as a function of the radial distance and the right panels
  show the same quantities as functions of the enclosed disk mass,
  $M_{\sm{encl}}$. Note that the initial rotation in the magnetic case
  is two time larger than in the two other cases (see also
  Sec.~\ref{sec:numerics})}
\label{fig:jz}
\eef

In Fig.~\ref{fig:jz} we compare the angular momentum
distribution and the rotational velocity for the three different
cases (Note that the overall enhancement of the angular momentum and
rotational velocity in the magnetized case is due to the initially
higher angular velocity). During this early stage of the collapse, 
the disk is still more massive than the central object(s) and is still
out of equilibrium. In all three cases the rotational velocity is
sub-Keplerian with 10\% to 60\% of the centrifugal
velocity. Apart from the general common features of the angular
momentum distribution of the three models there are important
differences. Both of the non-isothermal cases show a 'pile-up' of high
angular momentum gas at the shock fronts which sets the edge of the
disk. This can be seen in the lower left panel of Fig.~\ref{fig:jz}
where the high rotational velocity material accumulates at the shock
distance of $\mbox{few} \times 10 \, \AU$.  Inside the protodisk
(inside the shock fronts) the angular momentum is very efficiently
transported by the large bars and/or the magnetic field (see
Figs.~\ref{fig:disk_mag} and \ref{fig:disk_hydro}). Therefore, the
rotational velocity drops rapidly towards smaller radii in the hydro
and magnetized case. The isothermal case shows a much more evenly
distributed toroidal velocity when compared to the Keplerian
velocity. The reason here is the absence of shocks in the disk plane
which allows a continuous inflow of high angular momentum
gas. Additionally, the absence of a strong bar in the isothermal case
does not allow for efficient transport of angular momentum (see
Fig.~\ref{fig:disk_iso}). 

\bef
\showtwo{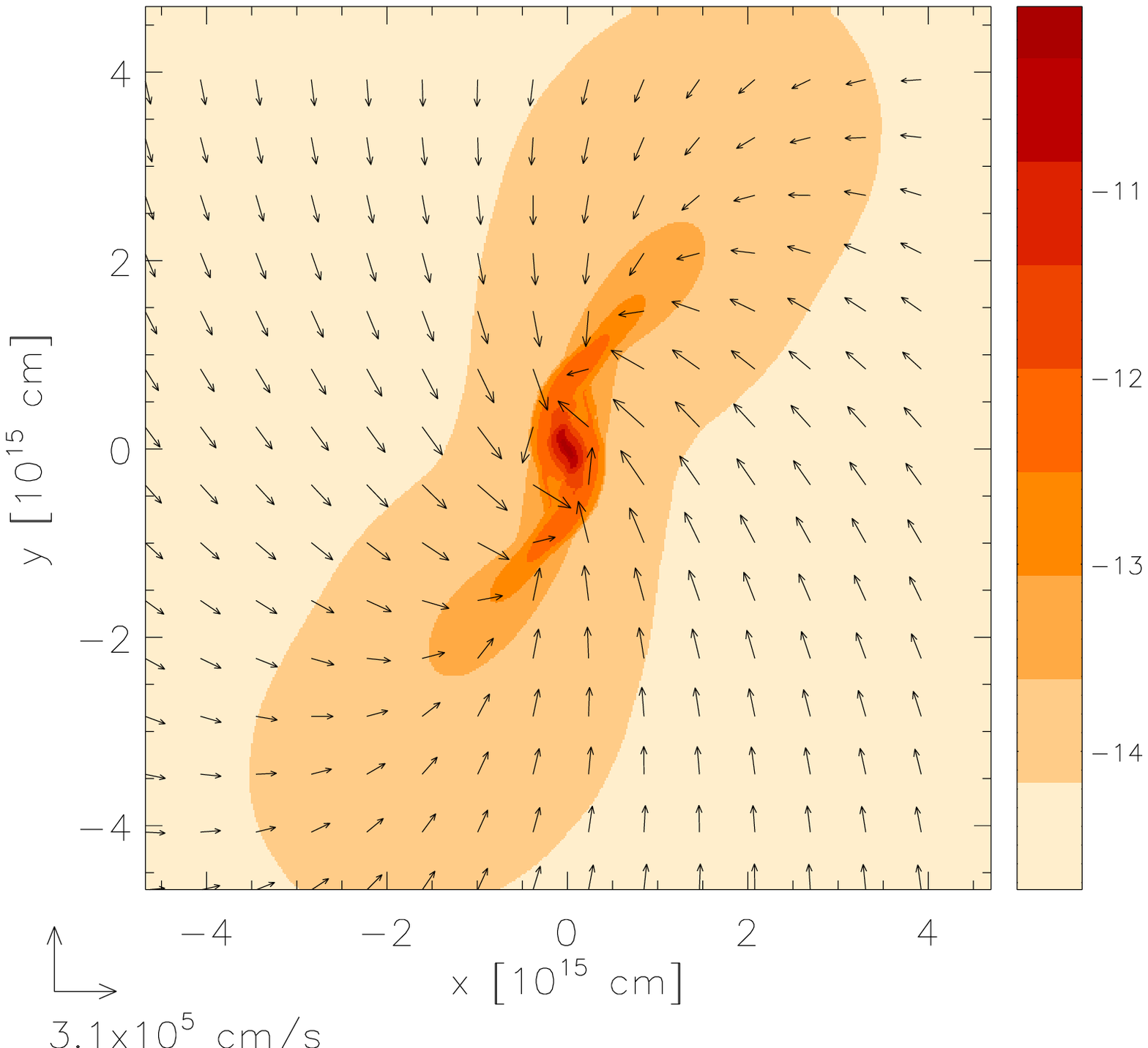}
	{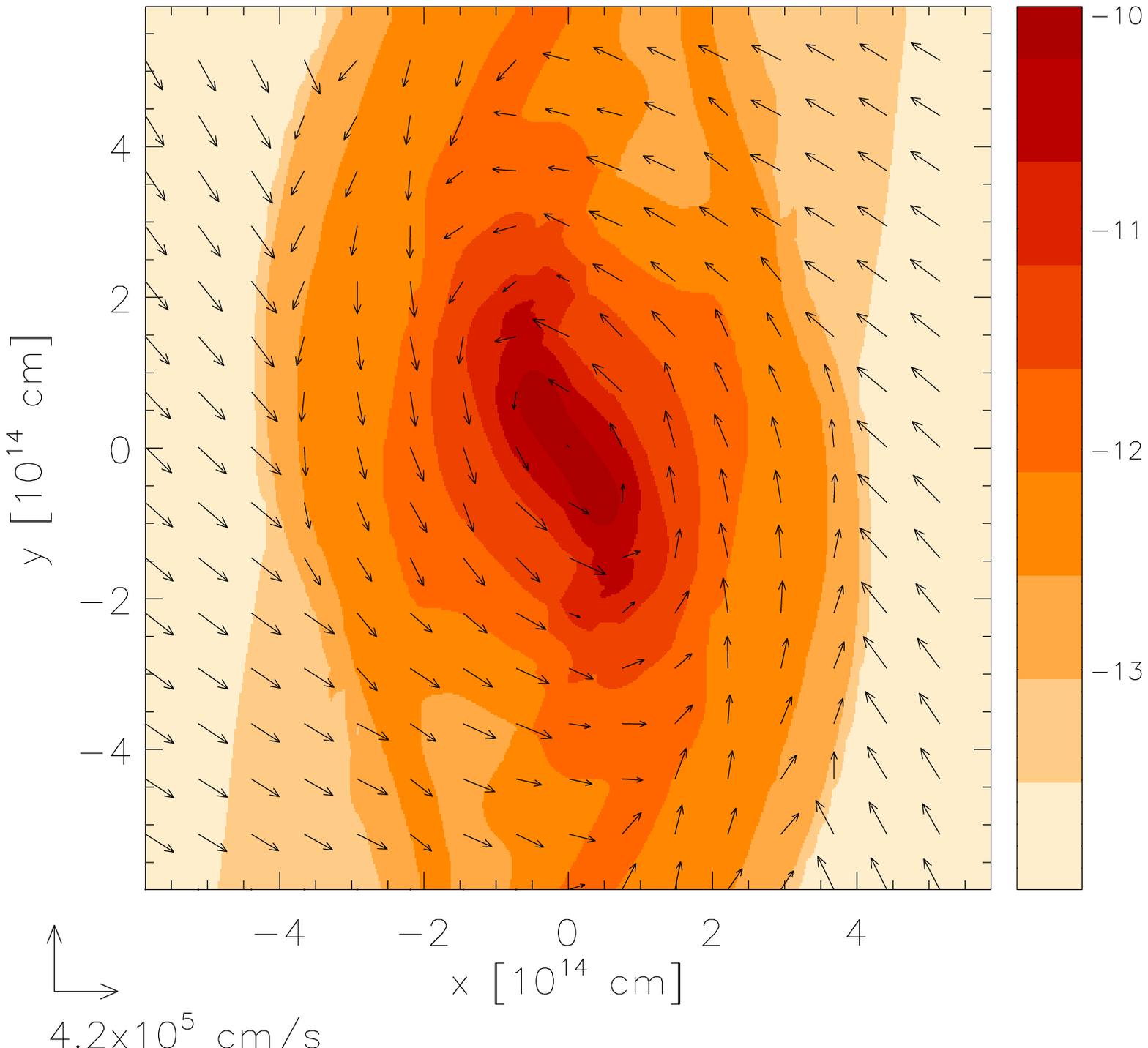}
\caption{Magnetized case: The panels show 2D slices through the disk
  plane at two different scales (large scales at right and small
  scales at the left). This snapshot is taken $1.4\times 10^4 \, \ys$
  into the collapse, corresponding to $\Sigma_{\sm{core}} \approx
  5\times 10^3 \, \g \, \cm^{-2}$ (see
  Fig.~\ref{fig:column_density}).}
\label{fig:disk_mag}
\eef

\bef
\showtwo{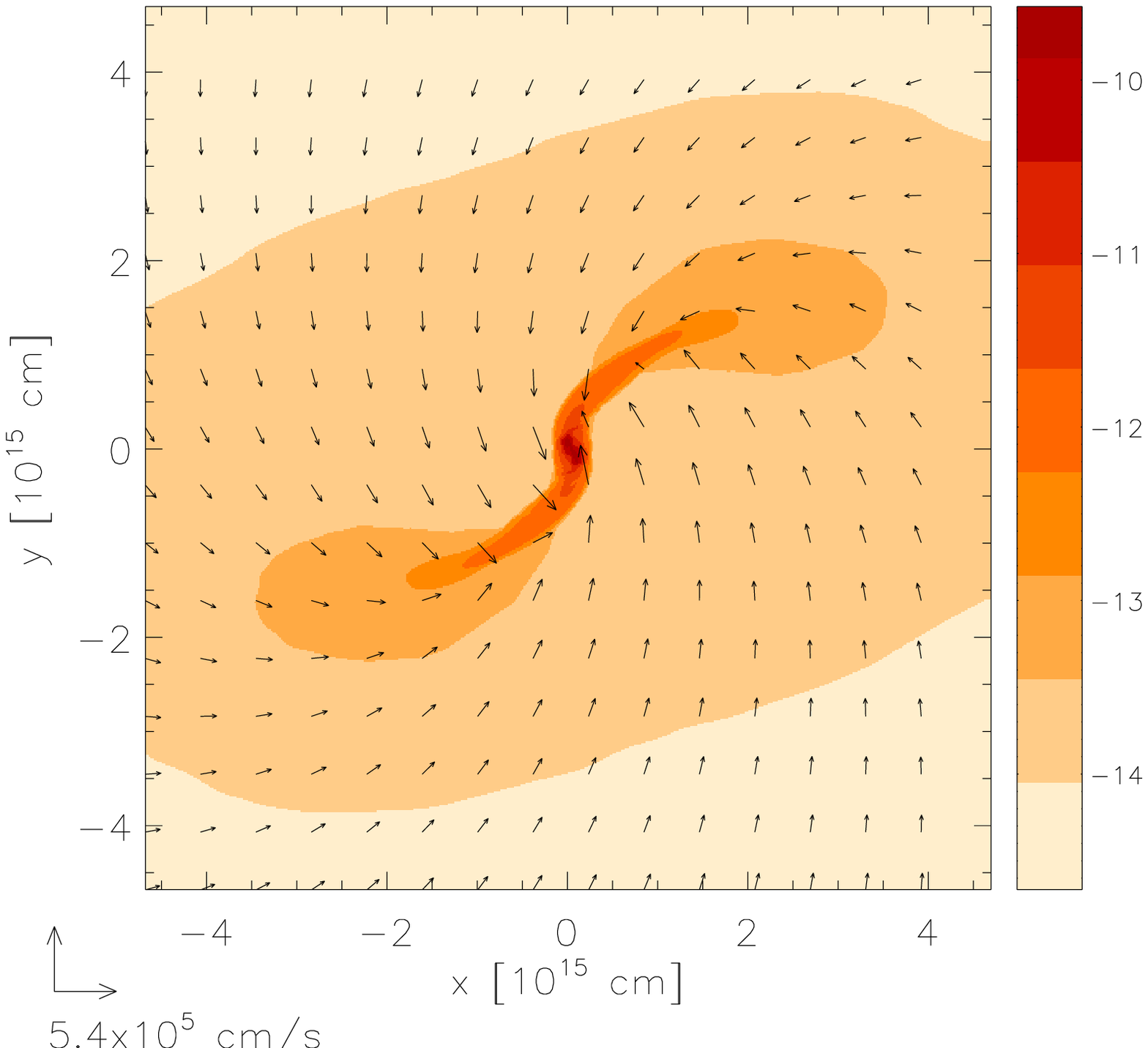}
	{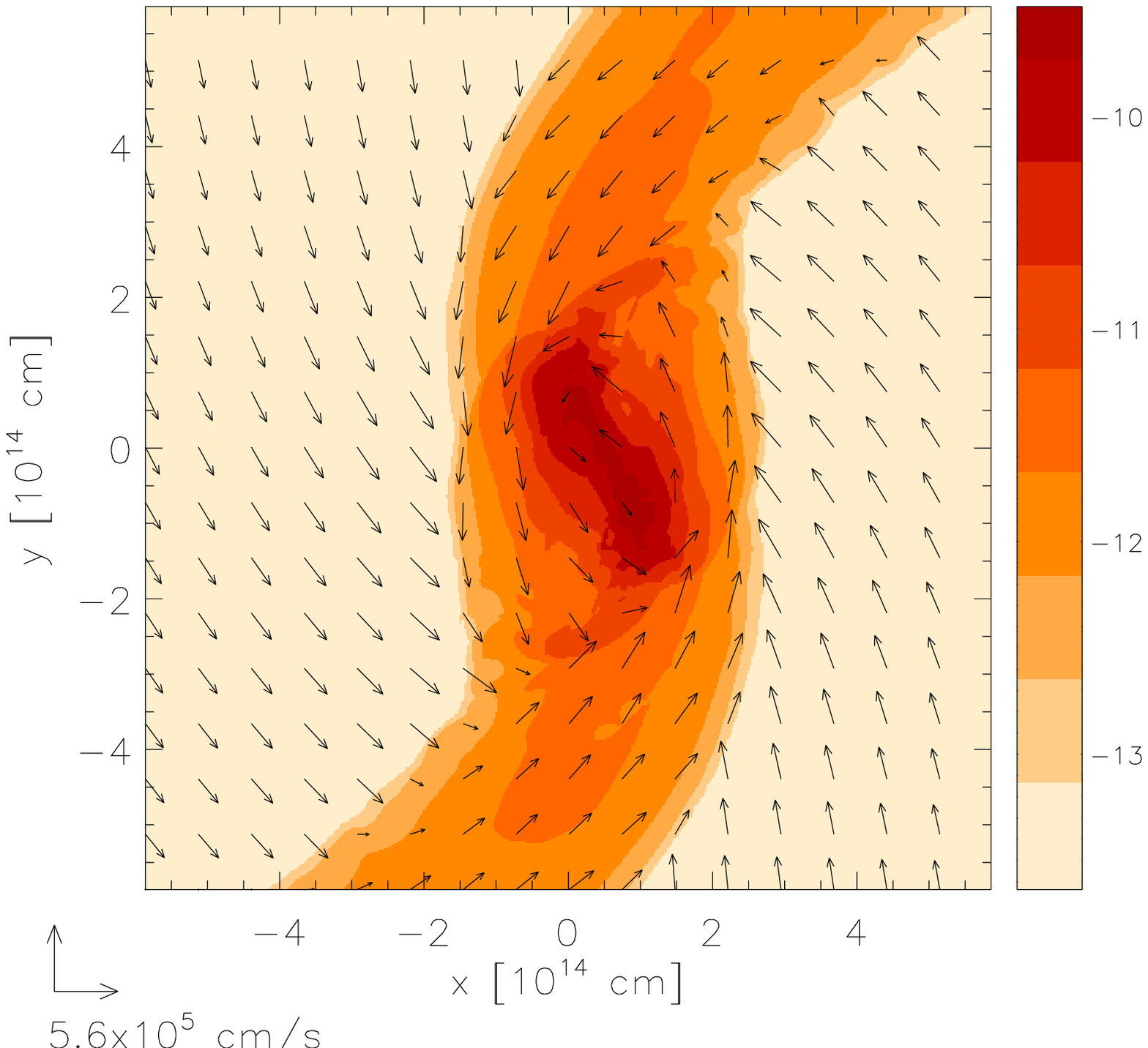}
\caption{The hydro case: the panels show 2D slices in the disk plane
  at two different scales (large scales at right and small scales at
  the left). This snapshot is taken $1.4\times 10^4 \, \ys$ into the
  collapse, corresponding to $\Sigma_{\sm{core}} \approx 5\times 10^3
  \, \g \, \cm^{-2}$ (see Fig.~\ref{fig:column_density}).}
\label{fig:disk_hydro}
\eef

\bef
\showtwo{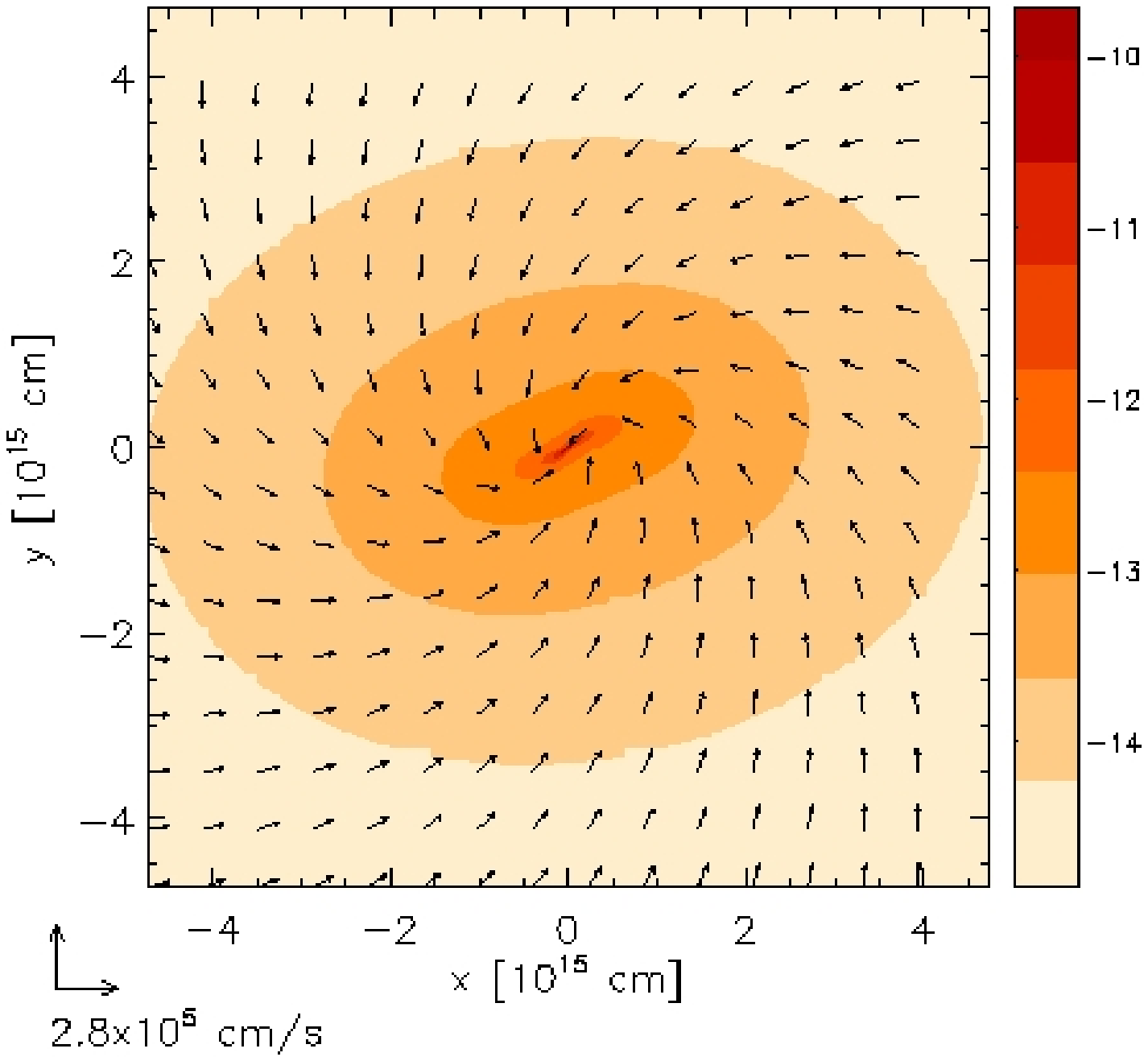}
	{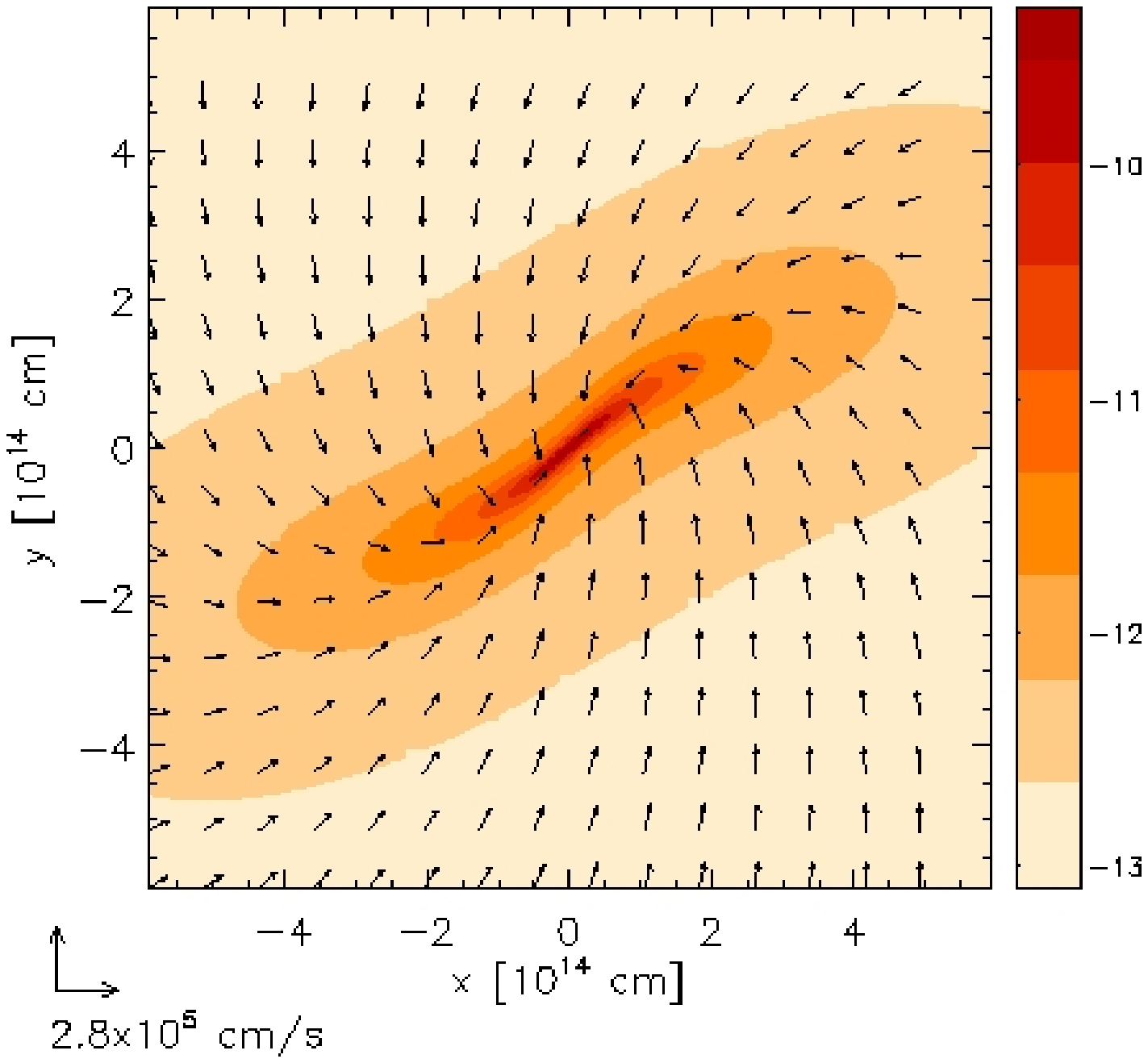}
%t_0 = t73 = 1.7968902e+14 sec (Sigma_core = 1.73)
\caption{The isothermal case: the panels show 2D slices in the disk plane
  at two different scales (large scales at right and small scales at
  the left). This snapshot is taken $10^4 \, \ys$ into the collapse,
  corresponding to $\Sigma_{\sm{core}} \approx 5\times 10^3 \, \g \,
  \cm^{-2}$ (see Fig.~\ref{fig:column_density}).}
\label{fig:disk_iso}
\eef

We investigate this further by presenting 2D snapshots of the disk
planes for the three different cases in Figs.~\ref{fig:disk_mag} --
\ref{fig:disk_iso} on a scale of $\sim 300 \, \AU$ and $\sim 40 \,
\AU$, respectively. Both the magnetized (Fig.~\ref{fig:disk_mag}) and
the hydro simulation (Fig.~\ref{fig:disk_hydro}) developed a large bar
with an extent of $\sim 250 \, \AU$. Comparing the small scale
structure of the magnetized and hydro case (right panel of
Figs.~\ref{fig:disk_mag} and \ref{fig:disk_hydro}) one sees that the
former case develops a larger, more rotationally supported
protodisk. This is due to the additional magnetic pressure and the
higher toroidal velocity in the magnetized case. Both cases accrete
efficiently gas through their bars/spiral arms onto the
protodisk. This can be seen particularly well in the right panel of
Fig.~\ref{fig:disk_hydro} where high density gas is streaming along
the spiral arms towards the central object. The isothermal case also
developed a bar but it is much smaller and less pronounced. Therefore,
the drop of the rotational velocity (measured in Keplerian velocity)
towards the disk center is only a factor of two (in the hydro case the
rotational velocity decreases by a factor of 6).

\section{Early outflows during massive star formation}
\label{sec:outflows}

\bef
\showtwo{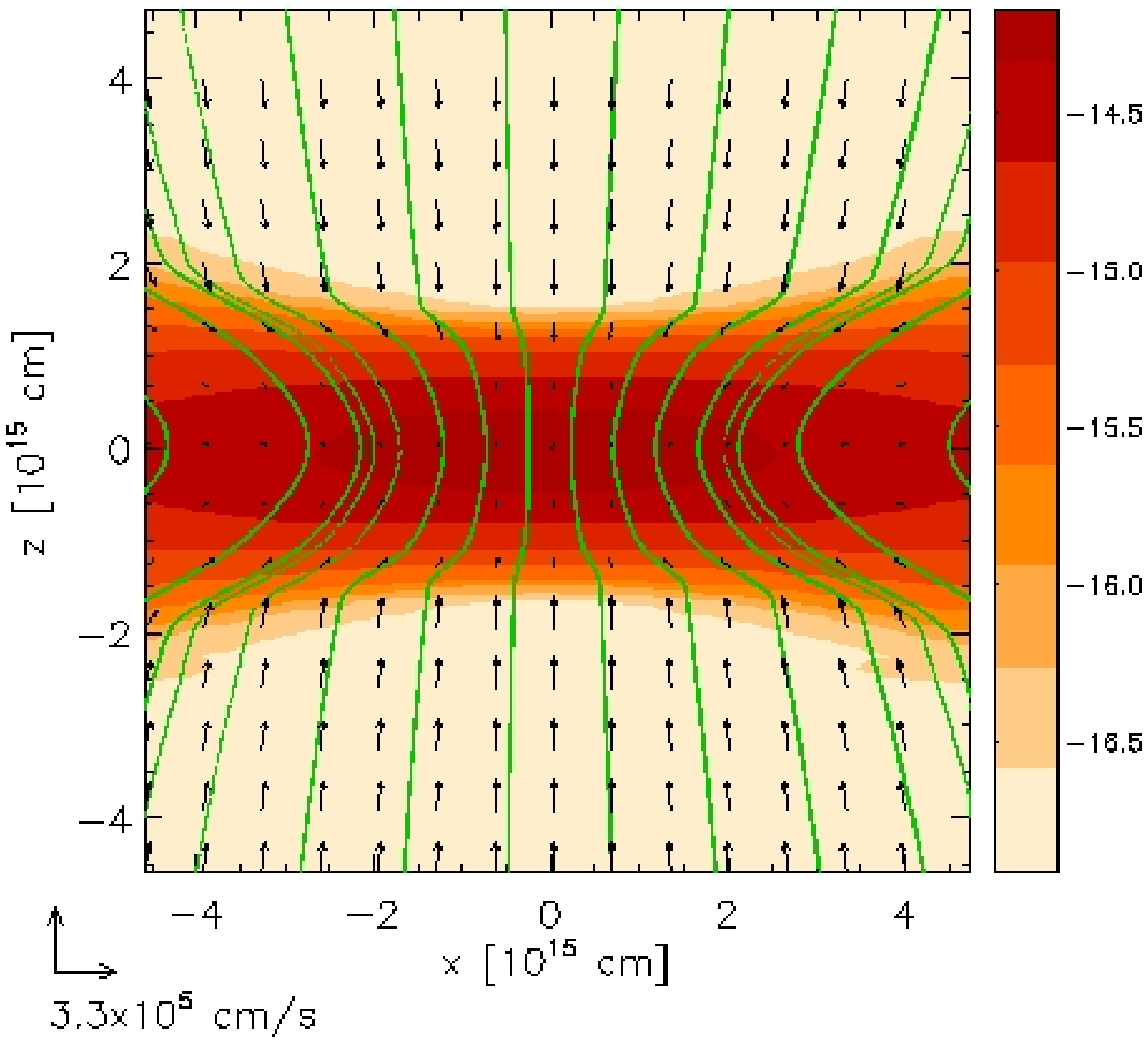}
	{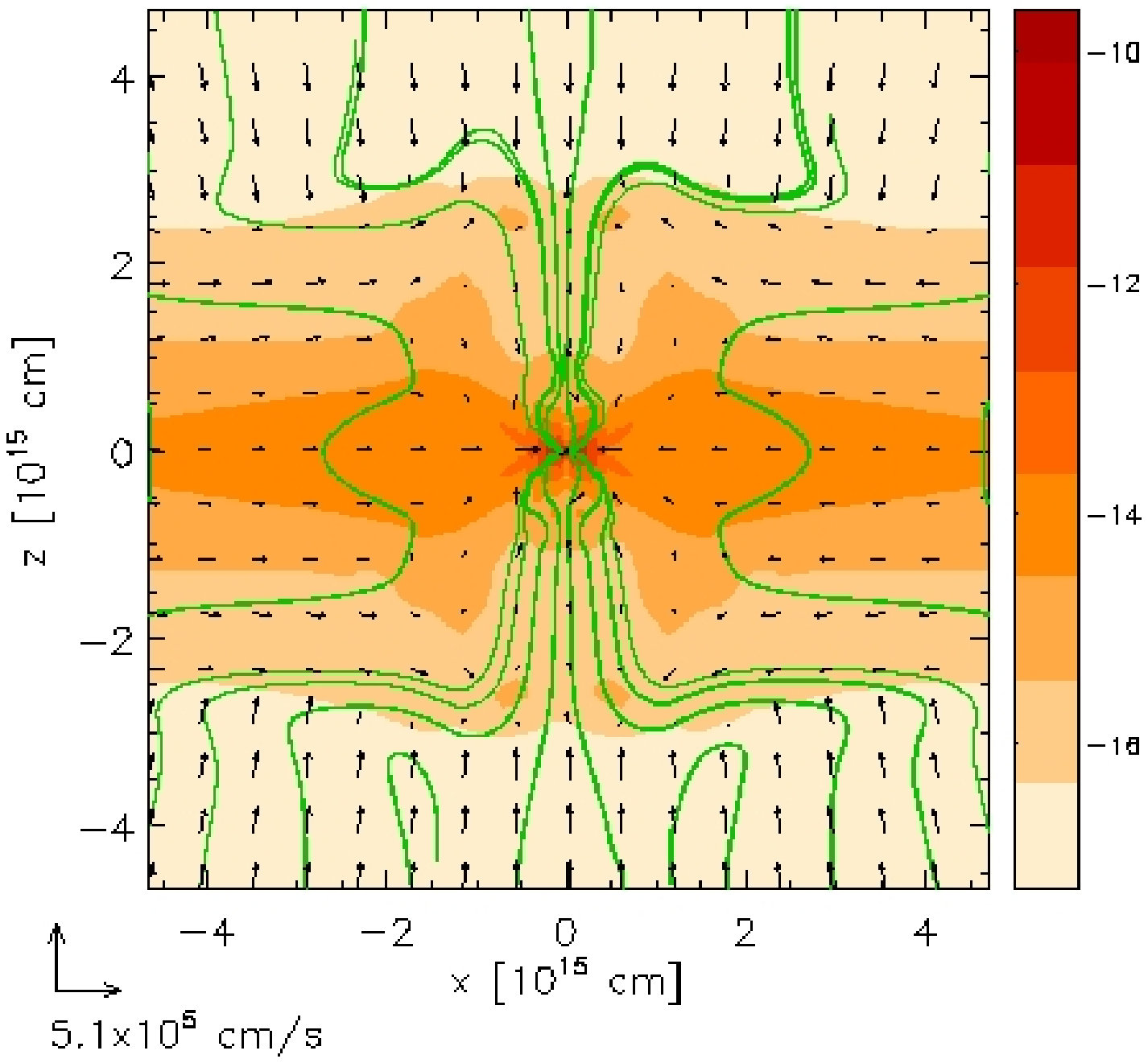}
% t143 = 1.6064849e+14 sec
% t320 = 1.6069792e+14 sec
\caption{Shows 2D snapshots perpendicular to the disk plane of the
  magnetized case at different times on large scales ($\sim 300 \,
  \AU$). The left panel shows the situation at $t = t_ 0 + 1.31\times
  10^4 \, \ys$ ($0.98 \, t_{\sm{ff}}$) into the collapse and the right
  panel shows the configuration $1566 \, \ys$ later after magnetic
  pressure inflates the disk and the outflow is launched (see also
  Fig.~\ref{fig:small_scale_outflow} for a close-up).}
\label{fig:large_scale_outflow}
\eef

\bef
\showtwo{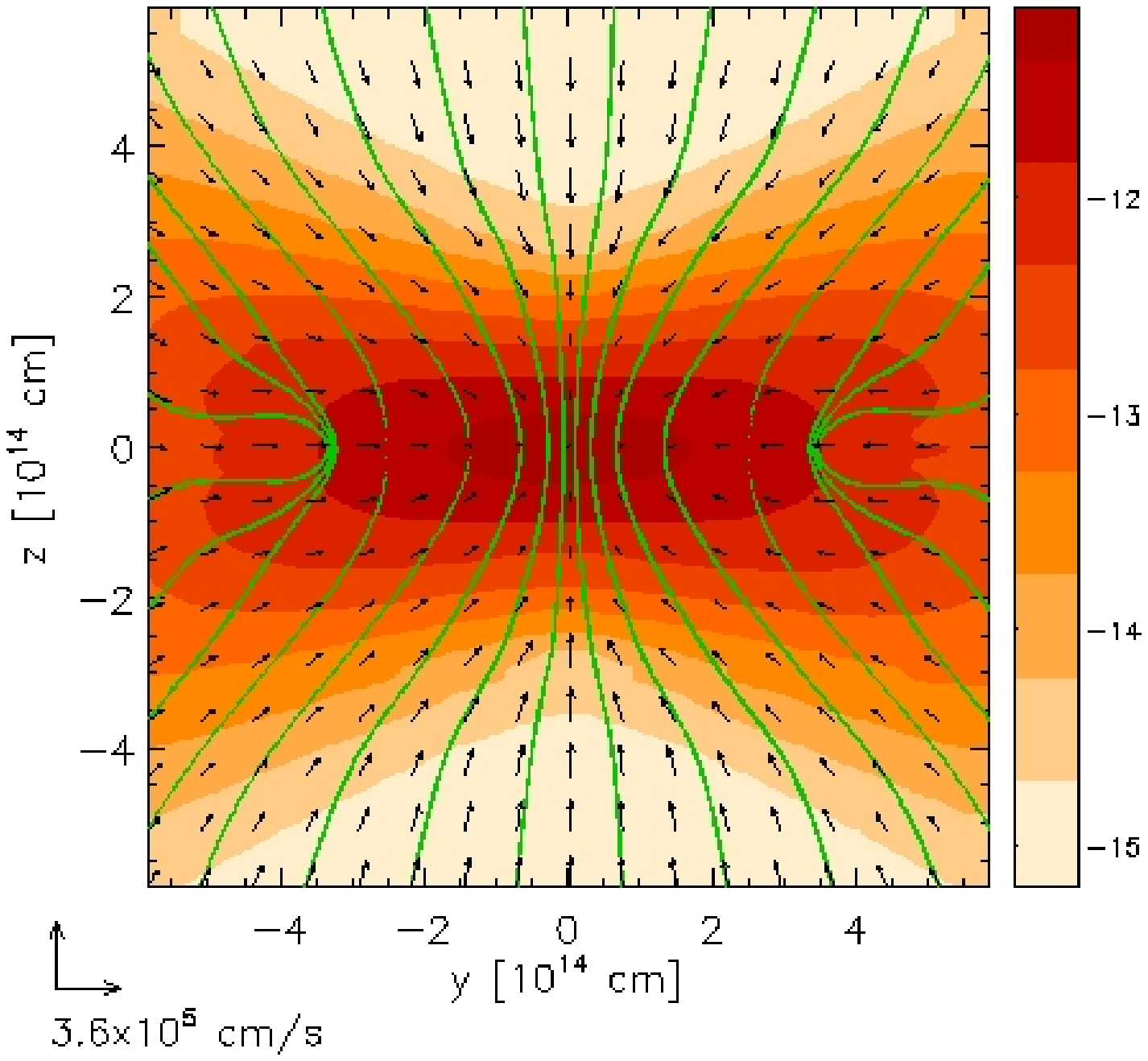}
	{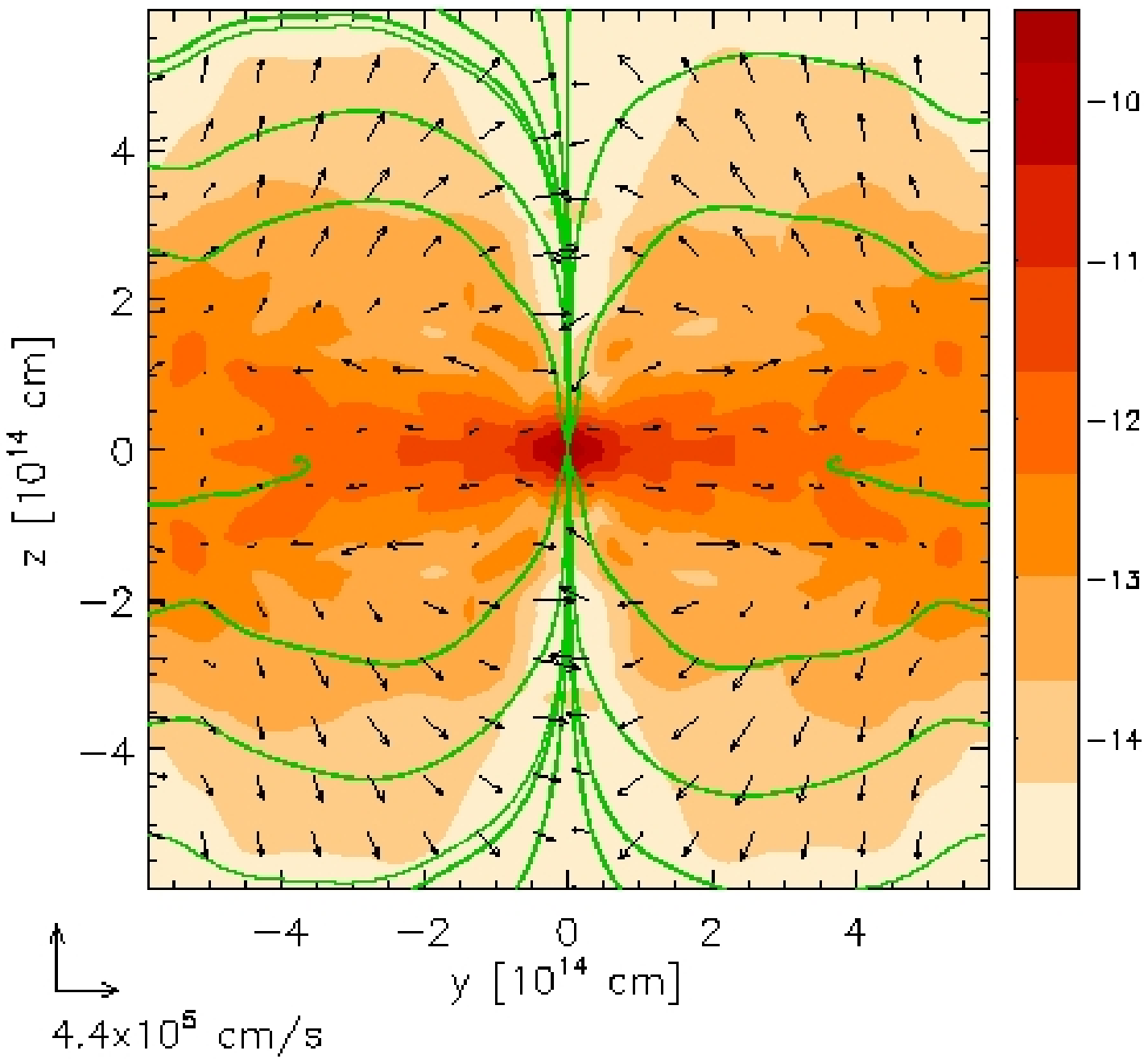}
% t143 = 1.6064849e+14 sec
% t180 = 1.6069197e+14 sec
% t240 = 1.6069663e+14 sec
% t320 = 1.6069792e+14 sec
\caption{Close-up 2D snapshots of the disk region (perpendicular to
  the disk in the $yz$-plane) in the magnetized case. The left panel shows the
  situation at $t = 1.45\times 10^4 \, \ys$ ($1.08 \, t_{\sm{ff}}$)
  into the collapse and before the flow reversal and right panel shows
  the configuration $188 \, \ys$ later when the outflow is clearly
  visible.}
\label{fig:small_scale_outflow}
\eef

Little is known observationally about the 
influence of outflows on the early assembly
of massive stars. On one hand, they could reduce the mass
accretion onto the massive (proto)star if the outflow carries a
substantial mass. On the other,  early outflows provide a natural
anisotropy of the accreting gas which results into low density
cavities. Such cavities are like funnels out of which the radiation from
the already active star can escape.  Without such a 
pressure-release valve, trapped radiative flux would halt the 
in-fall. \cite{Krumholz05} studied the effect of
outflow-funnels using a Monte Carlo radiative-transfer method which
shows that radiation pressure is greatly reduced by radiation escaping
through the outflow cavities.

Magnetic fields coupled to the protostellar disk can be the driving
power for such outflows. A variety of self-consistent simulations of
collapsing magnetized cloud cores show that outflows are launched if
the toroidal magnetic field pressure overcomes the gravitational force
\citep[][]{Tomisaka98, Tomisaka02, Matsumoto04, Machida04,
Banerjee06}. This early-type outflows can be understood in terms of a
growing magnetic tower \citep[][]{LyndenBell03}: The rotating
(proto)disk generates a strong toroidal field component by winding up
the threading field lines. The resulting magnetic pressure is in local
equilibrium with the gravitational force and the ram pressure of the
infalling material. But every further rotation increases the toroidal
field component thereby shifting the equilibrium location (which is
characterized by a shock) towards higher latitudes. The result is an
inflating magnetic bubble in which material is lifted off the disk.

Some of the highlights of this paper are shown in the following
figures, wherein 
we explicitly demonstrate the launch of the outflow.
We present 2D cuts
through our 3D data, which are shown as the images in 
Figs.~\ref{fig:large_scale_outflow} and \ref{fig:small_scale_outflow}.  
These show the collapse state at different times and physical scales. The time
sequence shown in Fig.~\ref{fig:large_scale_outflow} demonstrates the
situation shortly before the launching of the outflow (left panel) and
roughly $1600 \, \ys$ later (right panel). On this large scales ($\sim
300 \, \AU$) one can see the strong bending of the dragged-in magnetic
field lines (the toroidal component is not shown in the 2D slices) and
the inflated protodisk. Also clearly visible are the shock fronts
which separate the disk from the accreting environment. 

Close-ups
($\sim 20 \, \AU$) of the early stage of the outflow launching are
shown in Fig.~\ref{fig:small_scale_outflow}, where we zoom in 
by a factor of 10 in comparison with the previous figures. 
Here again the collapsing
stage shortly before the flow reversal is shown in the left panel and
the onset of the outflow (less than $200 \, \ys$ later) is shown in
the right panel. Due to the stronger magnetic field strength deeper 
in the gravitational potential well, the outflow velocity ($\sim 4 \, \km
\, \sec^{-1}$) is higher than on larger scales which powers a stronger
outflow.

It is also worth emphasizing that outflow only occurs when magnetic
fields are included in the simulations.  In this paper 
(as in our others), we have never
seen outflows associated with our purely hydrodynamic collapses.  
We conclude that purely hydrodynamic collapse misses a critical
ingredient in stellar formation - namely- the early launch of 
disk related outflows.  

\section{Summary and Conclusions}
\label{sec:conclusions}

In this work we simulated the 3D collapse of 
magnetized, massive molecular cloud cores which
are the progenitors of massive stars. 
We carried out this investigation at early stages
of the collapse, long before a massive star has even assembled
at the center of the resulting disk. Using three different 3D
collapse simulations (an isothermal case, a pure hydro case with
radiative cooling, and a magnetized case) of Bonnor-Ebert type cloud
cores we show that the mass accretion during the collapse phase is
much more efficient than calculated from the collapse of a singular
isothermal sphere (SIS). 

We find that the mass accretion rates can reach $\sim 10^2 \, c^3/G$
which is a factor of $10^2$ higher than the SIS case. This high mass
accretion is due to the supersonic infall velocity and the high
pressure at the edge of the collapsing core, where the pressure
increases with density as $P \propto \Sigma^2$. The mass accretion can
be estimated by $\dot{M} \sim v_r^3/G = \Ma^3 \, c^3 /G$ where the
Mach number, $\Ma$, can reach 3 or more with infall velocities ranging
from $1 - 1.5 \, \km \, \sec^{-1}$. Note that, due to inefficient
cooling at densities of $\Sigma \simge 10 \, \g \, \cm^{-2}$ the
temperature and therefore the sound speed increases which in turn
results in a higher physical mass accretion rates.  Therefore, mass
accretion rates of $10^{-3} \, \Msol \, \yr^{-1}$ and higher are
easily reached in the early stage of the assembly of massive
stars. This high accretion rates allow the formation of massive stars
in only $10^4 \, \ys$. A formation time scale of only $10^4 \, \ys$ is
short enough for the massive star to build up before its surrounding
disk is photo-evaporated from the UV environment emitted from the
surrounding OB association \citep[][]{Hollenbach00, Richling98}. This
allows also for an unified picture of star formation where low {\em
and} high mass star form the same way from the collapse of an
overdense cloud core. This unified theory is strongly supported by
recent observations of the assembly of massive stars through accretion
disks \citep{Chini04, Chini06} and from the detection of high
accretion rates onto a very young massive star \citep{Beltran06}.

Furthermore, we find that angular momentum is efficiently transfered
by bars/spiral arm and magnetic torque. The simulations with radiative
processes develop a larger bar-like structure than the isothermal case
and are less rotationally supported in the inner region of the
disk. The reduced specific angular momenta also enhance the accretion
rate as low angular momentum gas can be accreted more efficiently. In
particular we showed that the specific angular momentum decreases by a
factor of 2.8 within $14.7\times 10^3 \, \ys$ ($1.1 \, t_{\sm{ff}}$)
into the collapse. In all cases, an equilibrium disk did not yet
develop because at this early stage, 
the disk is still much more massive than the central
object. At this stage the rotational velocities are sub-Keplerian and
and range from $0.4 - 0.7 \, v_{\sm{Kep}}$ within the protodisk.

Outflows play an important role in this process by reducing the
radiation pressure which can escape through funnels carved by the
expelled gas \citep[][]{Krumholz05}. Here, we have shown that that outflows
are a natural consequence of the early stages of collapsing, massive
magnetized (and not purely hydrodynamic) cloud cores.  
This process is not restricted just to the
case of low mass star formation which we previously examined. Recent
observations indicate also early type outflows
\citep[e.g.,][]{Shepherd03} which might be associated with magnetic
fields. 

The critical issue that we have uncovered is 
that for both low (our previous work) and high mass star formation
(this work),
magnetic pressure from a wound-up toroidal field becomes eventually
strong enough to reverse the infalling gas and launch an outflow. We
find that the outflow has the shape of a hollow cylinder which outflow
speed reaches up to $5 \, \km \, \sec^{-1}$ at $20 \, \AU$ above and
below the disk midplane \citep[see e.g.][on details on magnetically
driven outflows]{Banerjee06}.

We conclude that this universality in the formation of 
disks and the early formation
of outflow cavities that are created by the jets that are
launched from their surfaces, 
implies that the radiation pressure from 
massive stars appears too late on the scene to substantially 
affect the 
assembly of a massive star by disk accretion.
We are pursuing more detailed computations of this picture
that will eventually include radiative feedback within
more complex, cluster forming environments.

\acknowledgments

We thank the anonymous referee for some useful remarks.
The FLASH  code was developed  in part by the  DOE-supported Alliances
Center for Astrophysical Thermonuclear Flashes (ASC) at the University
of Chicago. Our  simulations were carried out on  computer clusters of
the SHARCNET HPC Consortium. REP is supported by the Natural Sciences
and Engineering Research Council of Canada.

%\bibliographystyle{apj}
%\bibliography{astro}

\end{document}